\PassOptionsToPackage{unicode}{hyperref}
\PassOptionsToPackage{hyphens}{url}
\PassOptionsToPackage{dvipsnames,svgnames,x11names}{xcolor}
\documentclass[
  11pt,
  a4paper,
]{article}
\usepackage{xcolor}
\usepackage[a4paper,textheight=24cm,textwidth=15.5cm]{geometry}
\usepackage{amsmath,amssymb}
\usepackage{iftex}
\ifPDFTeX
  \usepackage[T1]{fontenc}
  \usepackage[utf8]{inputenc}
  \usepackage{textcomp} % provide euro and other symbols
\else % if luatex or xetex
  \usepackage{unicode-math} % this also loads fontspec
  \defaultfontfeatures{Scale=MatchLowercase}
  \defaultfontfeatures[\rmfamily]{Ligatures=TeX,Scale=1}
\fi
\usepackage[]{libertinus}
\ifPDFTeX\else
\fi
\IfFileExists{upquote.sty}{\usepackage{upquote}}{}
\IfFileExists{microtype.sty}{% use microtype if available
  \usepackage[]{microtype}
  \UseMicrotypeSet[protrusion]{basicmath} % disable protrusion for tt fonts
}{}
\makeatletter
\@ifundefined{KOMAClassName}{% if non-KOMA class
  \IfFileExists{parskip.sty}{%
    \usepackage{parskip}
  }{% else
    \setlength{\parindent}{0pt}
    \setlength{\parskip}{6pt plus 2pt minus 1pt}}
}{% if KOMA class
  \KOMAoptions{parskip=half}}
\makeatother
\makeatletter
\ifx\paragraph\undefined\else
  \let\oldparagraph\paragraph
  \renewcommand{\paragraph}{
    \@ifstar
      \xxxParagraphStar
      \xxxParagraphNoStar
  }
  \newcommand{\xxxParagraphStar}[1]{\oldparagraph*{#1}\mbox{}}
  \newcommand{\xxxParagraphNoStar}[1]{\oldparagraph{#1}\mbox{}}
\fi
\ifx\subparagraph\undefined\else
  \let\oldsubparagraph\subparagraph
  \renewcommand{\subparagraph}{
    \@ifstar
      \xxxSubParagraphStar
      \xxxSubParagraphNoStar
  }
  \newcommand{\xxxSubParagraphStar}[1]{\oldsubparagraph*{#1}\mbox{}}
  \newcommand{\xxxSubParagraphNoStar}[1]{\oldsubparagraph{#1}\mbox{}}
\fi
\makeatother

\usepackage{longtable,booktabs,array}
\usepackage{calc} % for calculating minipage widths
\usepackage{etoolbox}
\makeatletter
\patchcmd\longtable{\par}{\if@noskipsec\mbox{}\fi\par}{}{}
\makeatother
\IfFileExists{footnotehyper.sty}{\usepackage{footnotehyper}}{\usepackage{footnote}}
\makesavenoteenv{longtable}
\usepackage{graphicx}
\usepackage{fvextra}
\DefineVerbatimEnvironment{Routput}{Verbatim}{breaklines=true,breakanywhere=true}
\makeatletter
\newsavebox\pandoc@box
\newcommand*\pandocbounded[1]{% scales image to fit in text height/width
  \sbox\pandoc@box{#1}%
  \Gscale@div\@tempa{\textheight}{\dimexpr\ht\pandoc@box+\dp\pandoc@box\relax}%
  \Gscale@div\@tempb{\linewidth}{\wd\pandoc@box}%
  \ifdim\@tempb\p@<\@tempa\p@\let\@tempa\@tempb\fi% select the smaller of both
  \ifdim\@tempa\p@<\p@\scalebox{\@tempa}{\usebox\pandoc@box}%
  \else\usebox{\pandoc@box}%
  \fi%
}
\def\fps@figure{htbp}
\makeatother

\NewDocumentCommand\citeproctext{}{}

\makeatletter
 \let\@cite@ofmt\@firstofone
 \def\@biblabel#1{}
 \def\@cite#1#2{{#1\if@tempswa , #2\fi}}
\makeatother
\newlength{\cslhangindent}
\newlength{\csllabelwidth}
\newenvironment{CSLReferences}[2] % #1 hanging-indent, #2 entry-spacing
 {\begin{list}{}{%
  \setlength{\itemindent}{0pt}
  \setlength{\leftmargin}{0pt}
  \setlength{\parsep}{0pt}
  \ifodd #1
   \setlength{\leftmargin}{\cslhangindent}
   \setlength{\itemindent}{-1\cslhangindent}
  \fi
  \setlength{\itemsep}{#2\baselineskip}}}
 {\end{list}}
\usepackage{calc}

\providecommand{\tightlist}{%
  \setlength{\itemsep}{0pt}\setlength{\parskip}{0pt}}

\usepackage{tikz}
\usepackage{xcolor}
\usepackage{tabularx}
\usepackage{orcidlink}
\usepackage{lineno}
\usepackage{libertinus}
\usepackage{draftwatermark}

\makeatletter
\@ifpackageloaded{caption}{}{\usepackage{caption}}
\AtBeginDocument{%
\ifdefined\contentsname
  \renewcommand*\contentsname{Table of contents}
\else
  \newcommand\contentsname{Table of contents}
\fi
\ifdefined\listfigurename
  \renewcommand*\listfigurename{List of Figures}
\else
  \newcommand\listfigurename{List of Figures}
\fi
\ifdefined\listtablename
  \renewcommand*\listtablename{List of Tables}
\else
  \newcommand\listtablename{List of Tables}
\fi
\ifdefined\figurename
  \renewcommand*\figurename{Figure}
\else
  \newcommand\figurename{Figure}
\fi
\ifdefined\tablename
  \renewcommand*\tablename{Table}
\else
  \newcommand\tablename{Table}
\fi
}
\@ifpackageloaded{float}{}{\usepackage{float}}
\floatstyle{ruled}
\@ifundefined{c@chapter}{\newfloat{codelisting}{h}{lop}}{\newfloat{codelisting}{h}{lop}[chapter]}
\floatname{codelisting}{Listing}

\makeatother
\makeatletter
\@ifpackageloaded{caption}{}{\usepackage{caption}}
\@ifpackageloaded{subcaption}{}{\usepackage{subcaption}}
\makeatother
\makeatletter
\@ifpackageloaded{algorithm}{}{\usepackage{algorithm}}
\makeatother
\makeatletter
\@ifpackageloaded{algpseudocode}{}{\usepackage{algpseudocode}}
\makeatother
\makeatletter
\@ifpackageloaded{caption}{}{\usepackage{caption}}
\makeatother
    \makeatletter
    \newcommand\fs@nocaption{
      \def\@fs@cfont{\bfseries}
      \let\@fs@capt\floatc@plain
      \def\@fs@pre{}%
      \def\@fs@post{\kern2pt\hrule}%
      \def\@fs@mid{\hrule\kern2pt}%
      \let\@fs@iftopcapt\iftrue}
    \makeatother
  
\usepackage{bookmark}
\IfFileExists{xurl.sty}{\usepackage{xurl}}{} % add URL line breaks if available
\hypersetup{
  pdftitle={Detecting adverse high-order drug combinations from individual case safety reports using computational statistics on disproportionality measures},
  pdfauthor={Jules Bangard; Einar Holsbø; Kristian Svendsen; Vittorio Perduca; Étienne Birmelé},
  pdfkeywords={Disproportionality
Analysis, Drug-Combinations, Drug-Interactions, Genetic
Algorithm, Markov Chain Monte Carlo, Pharmacovigilance, Spontaneous
Reporting Systems},
  colorlinks=true,
  linkcolor={blue},
  filecolor={Maroon},
  citecolor={Blue},
  urlcolor={Blue},
  pdfcreator={LaTeX via pandoc}}

\title{Detecting adverse high-order drug combinations from individual
case safety reports using computational statistics on disproportionality
measures}
\author{Jules Bangard \and Einar Holsbø \and Kristian
Svendsen \and Vittorio Perduca \and Étienne Birmelé}
\date{2026-07-14}
\begin{document}
\definecolor{computo-blue}{HTML}{034E79}

\begin{tikzpicture}[remember picture,overlay]
\fill[computo-blue]
  (current page.north west) -- (current page.north east) --
  ([yshift=-5cm]current page.north east|-current page.north east) --
  ([yshift=-5cm]current page.north west|-current page.north west) -- cycle;
\node[anchor=north west, xshift=.75cm,
  yshift=-.75cm] at (current page.north west) {\includegraphics[height=3cm]{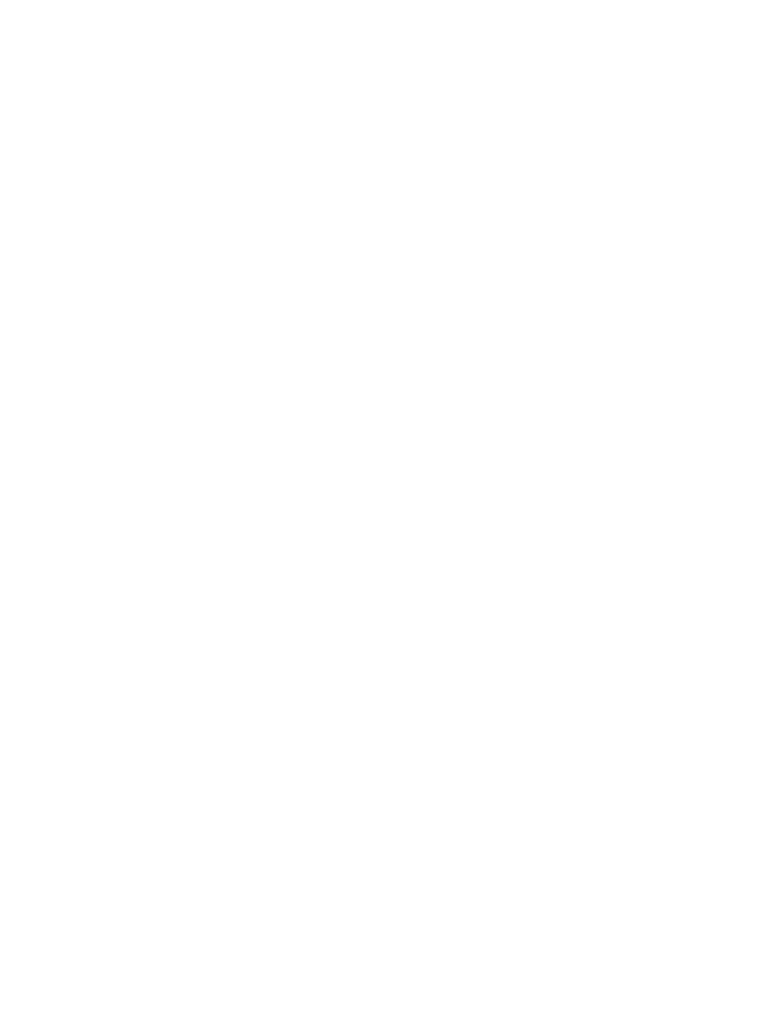}};
\node[font=\sffamily\bfseries\color{white},anchor=north west, xshift=.75cm,
  yshift=-4.25cm] at (current page.north
  west) {\fontsize{10}{12}\selectfont ISSN 2824-7795};
\node[font=\sffamily\bfseries\color{white},anchor=west,
  xshift=4.25cm,yshift=-2.75cm] at (current page.north west)
  {\begin{minipage}{15cm}
    \fontsize{25}{30}\selectfont
    Detecting adverse high-order drug combinations from individual case
    safety reports using computational statistics on disproportionality
    measures
    \vspace{.5cm}
    \\
    \fontsize{15}{18}\selectfont
    
  \end{minipage}};
\end{tikzpicture}

\vspace*{2.5cm}
\begin{center}
          Jules
Bangard~\orcidlink{0009-0007-4670-7860}\footnote{Corresponding author: \href{mailto:jules.bangard@math.unistra.fr}{jules.bangard@math.unistra.fr}}\quad
             Institut de Recherche Mathématique Avancée, UMR 7501
Université de Strasbourg et CNRS 7 rue René-Descartes, 67000 Strasbourg,
France\\
                 Einar Holsbø~\orcidlink{0000-0002-9728-2088}\quad
             Faculty of Science and Technology, UiT-The Arctic
University of Norway, PO, Box 6050 Stakkevollan, N-9037 Tromsø, Norway\\
                 Kristian Svendsen~\orcidlink{0000-0003-3481-3539}\quad
             Faculty of Health Sciences, UiT the Arctic University of
Norway, Tromsø, Norway\\
                 Vittorio Perduca~\orcidlink{0000-0003-0339-0473}\quad
             CNRS, MAP5, Université Paris Cité, F-75006 Paris, France\\
                 Étienne Birmelé~\orcidlink{0000-0002-6996-4014}\quad
             Institut de Recherche Mathématique Avancée, UMR 7501
Université de Strasbourg et CNRS 7 rue René-Descartes, 67000 Strasbourg,
France\\
           
  \bigskip
  
  Date published: 2026-07-14 \quad Last modified: 2026-07-14
\end{center}
      
\bigskip
\begin{abstract}
Adverse drug reactions linked to the intake of drug combinations are a
critical concern in pharmacovigilance, particularly as the controlled
environment of clinical trials often lacks the scale and diversity to
detect rare events involving multiple medications. While spontaneous
reporting systems provide the necessary breadth for post-market
surveillance, identifying overrepresented drug cocktails within such
large-scale data remains a significant computational challenge. This
study introduces a computational framework for the detection of drug
cocktails associated with adverse events, leveraging disproportionality
analysis on individual case safety reports. By integrating the
Anatomical Therapeutic Chemical classification, the framework extends
beyond individual drugs to capture hierarchical pharmacological
relationships, enabling exploration of the space of drug combinations
beyond pairwise analysis. To address biases inherent in existing
disproportionality measures, we employ a hypergeometric risk metric,
while a Markov Chain Monte Carlo algorithm provides robust empirical
p-value estimation for the risk associated with cocktails. A genetic
algorithm further facilitates efficient identification of high-risk drug
cocktails. A post-treatment step based on penalized logistic regression
allows distinguishing true pharmacological interactions from combined
individual effects for cocktails of any size. Validation on synthetic
and FDA Adverse Event Reporting System data demonstrates the method's
efficacy in detecting established drugs and drug combinations associated
with myopathy-related adverse events. Implemented as an R package, this
framework offers a reproducible, scalable tool for post-market drug
safety surveillance.
\end{abstract}

\noindent%
\vfill

\DraftwatermarkOptions{stamp=false}

\floatname{algorithm}{Algorithm}

\renewcommand*\contentsname{Contents}
{
\hypersetup{linkcolor=}
\setcounter{tocdepth}{3}
\tableofcontents
}

\section{Introduction}\label{sec-intro}

There are inherent limitations of clinical trials (RCTs) done before a
medication is authorised in terms of cost, surveillance time, size and a
lack of diversity of patients included (e.g.~that pregnant women,
children and elderly often are not included) (Sanson-Fisher et al.
2007). Another issue is that RCTs are limited in their ability to detect
rare adverse drug reactions (ADRs) and potential drug interactions
(DDIs) since they usually only include a few thousand participants at
most. Moreover, the concurrent use of multiple medications is often an
exclusion criterion in RCTs, leaving gaps in the assessment of
polypharmacy risks. As a result, tested drug combinations tend to be
limited to those where physicians and pharmacologists have existing
clinical experience or hypotheses about potential interactions (Heijden
et al. 2002). This means that monitoring of the safety of medications
and the potential for DDIs after marketing is essential. This monitoring
is often referred to as post-authorization pharmacovigilance.

Elderly individuals are commonly prescribed multiple medications due to
the increasing prevalence of comorbidities associated with aging. One
study suggested that patients aged \(75\) or older in Austria took on
average \(7.5\) drugs (Schuler et al. 2008). Among this demographic,
\(10\%\) of hospitalizations were attributed to ADRs, and \(18.7\%\) of
these ADRs were caused by a combination of more than one drug, a
drug-drug interaction. More broadly, a meta-analysis attributed
approximately \(22.2\%\) of hospitalizations caused by adverse drug
reactions to DDIs (Dechanont et al. 2014).

Disproportionality analysis (DA) encompasses widely used methods in
pharmacovigilance to detect ADRs by assessing the frequency of adverse
events (AE) associated with specific drug consumption relative to what
would be expected. Most established disproportionality methods like the
proportional reporting ratio (PRR) (Evans, Waller, and Davis 2001),
reporting odds ratio (ROR) (Puijenbroek et al. 2002), Bayesian
Confidence Propagation Neural Network (BCPNN) (Bate et al. 1998), and
the Gamma Poisson Shrinker (GPS) (DuMouchel 1999) allow for the
identification of signals that warrant further investigation in the
context of single drug assessment. The tree-based scan statistic
(Kulldorff, Fang, and Walsh 2003; Heo et al. 2024) takes advantage of
the tree structure of the Anatomical Therapeutic Chemical (ATC) drug
classification in order to allow selection of a drug family rather than
a single drug, a family corresponding to a subtree of the ATC tree.
Likelihood-ratio tests then allow selection of the most relevant family.

In order not to overinterpret high proportions in case of rare drug
cocktails, it is possible to derive a score from an independence test on
the contingency table counting the presence or absence of the adverse
event depending on the consumption of the drug cocktail. (Gosho et al.
2017) use a chi-square statistic to do so, while (Ahmed et al. 2010)
rather use the Fisher exact test which has the advantage of being
non-asymptotic.

Fewer DA methods are available to use for multiple drug consumption,
such as the \(\Omega\) shrinkage method (Norén et al. 2008). This method
allows the use of a DA measure in order to detect sets of the type
\emph{Drug-Drug-Adverse Event} and stands as the standard measure when
it comes to detecting the interactions of two drugs. Adaptations of the
PRR for a single drug have been proposed like the Concomitant Signal
Score (CSS) (Noguchi et al. 2020) and the PRR adaptation for DDIs (Wang
et al. 2020).

Other methods than DA measures exist in order to detect DDIs. Among
them, well known methods encompass logistic models (Van Puijenbroek et
al. 2000, 1999) and association rules methods (Noguchi et al. 2018;
Ibrahim et al. 2016). One can refer to multiple reviews for a more
detailed overview of existing methods (Ibrahim et al. 2021; Hauben
2023).

Computational pharmacovigilance methods are typically applied to
Spontaneous Reporting Systems (SRS), which aggregate Individual Case
Safety Reports (ICSRs)---real-world data submitted by healthcare
professionals, manufacturers, and patients. Such reports contain
information about intake of medications for each patient as well as
their experienced AE. While SRS databases, such as the FDA Adverse Event
Reporting System (FAERS), provide the scale necessary to detect rare
events, they possess inherent limitations. These include
under-reporting, where only a fraction of adverse events are captured
(Bate and Evans 2009), and various reporting biases, such as the
notoriety bias (Pariente et al. 2007). Crucially, the spontaneity of
these reports often results in an incomplete recording of all drugs
taken by a patient. Furthermore, because the exact timing of drug
administration is frequently omitted, it is difficult to distinguish
between simultaneous intake and medications taken at different times
within the same reporting window. This temporal ambiguity potentially
limits the precision of interaction analyses, as the observed cocktail
may represent a sequence of exposures rather than a concurrent
pharmacological event. Consequently, because the total number of
patients exposed to a drug is unknown, these databases cannot provide
true incidence rates, necessitating the use of disproportionality
analysis to identify safety signals (Almenoff et al. 2005).

\medskip

While previous work showed that further practical experiences should be
of interest for high-order drug interaction testing (Tekin et al. 2018),
it is hard to know which exact combination one should test and easy to
miss an important combination in the overwhelming space of high-order
drug cocktails. The present article proposes a novel method to identify
drug cocktails that are overrepresented in adverse event reports, which
constitute candidates for further interaction assessment. It requires
the choice of a scoring function able to measure, for a given AE, the
importance of the disproportionality for any medication set, including
single medications. It then takes advantage of the tree structure of the
ATC classification to explore the space of medication sets in two ways:
a genetic algorithm looking for the sets of high score, and a MCMC
algorithm learning the distribution of the scores for a given cocktail
size in order to evaluate how extreme a high score is. A post-treatment
step using penalized logistic regression then allows to assess whether
an identified cocktail reflects a true pharmacological interaction or
merely the combined effect of its components. The approach is also
original with respect to the fact that a medication denotes either a
drug or a drug family in the sense of an internal node of the ATC
classification.

The developed algorithms, available in the corresponding R package
\textbf{emcAdr} (Bangard 2025), aim to guide pharmacology researchers
toward drug combinations that might correspond to drug interactions and
require more rigorous monitoring.

\section{Methods}\label{methods}

\subsection{ATC Tree and Cocktail Definition}\label{sec-cocktail}

Multiple classifications of drug active ingredients exist, one being the
ATC classification. This system is structured hierarchically and can be
represented as a tree with five levels containing, at the time of this
study, \(6809\) nodes. The leaves of this tree are the active
ingredients, while the first level consists of nodes representing organs
or systems affected by the descendant active ingredients. The remaining
nodes correspond to therapeutic or pharmacological families.

By applying a Depth-First Search algorithm to enumerate each node of the
ATC tree, we can define a drug cocktail as a set of integers
corresponding to specific nodes in the tree. More formally, a cocktail
\(C\) of size \(k\geq 1\) is defined by
\(C = (x_1, \dots, x_k) \in \Delta^{k}\) where \(k \in \mathbb{N}\) ,
and \(\Delta\) represents the set of numbered nodes of the ATC tree
\(T\). Figure~\ref{fig-mut} shows examples in a simplified tree, the
green nodes denoting the considered cocktail. Note that for convenience,
we also refer to single drugs as cocktails.

Considered cocktails can include internal nodes of the tree (thus
representing families of active ingredients), allowing for the detection
of more general signals. For example, paracetamol might send a weak
signal, while if we move up the tree, analgesics as a whole could
represent a stronger and more general signal. Therefore, all patients
taking at least one drug from this drug family will be considered in the
computation of the risk. That notion is equivalent to the notion of cut
in the tree-scan method (Kulldorff, Fang, and Walsh 2003).

\subsection{Cocktail Risk Characterization}\label{sec-risk}

In the field of pharmacology, accurately characterizing the risk
associated with drug administration is a complex task. The aim of the
developed method is to search the space of cocktails to maximize a score
indicating if individuals taking a given drug combination have a higher
risk of an AE. While the algorithm is designed to be flexible and can be
run with any scoring function, we focus on a comparison of established
metrics to justify our selection (see Section~\ref{sec-score_comp}).

\subsubsection{Scoring Functions in
Pharmacovigilance}\label{sec-risk-definition}

To evaluate the disproportionality of drug cocktails presence for a
particular AE, several statistical measures have been proposed in the
literature:

\begin{description}
\tightlist
\item[Proportional Reporting Ratio (RR)]
A standard measure defined as the ratio of the probability of an AE
occurring in the group exposed to the cocktail versus the non-exposed
group. Its application in signal generation from spontaneous reports was
notably discussed by (Evans, Waller, and Davis 2001).
\item[PRR for Drug-Drug Interactions (PRR)]
Unlike the single-drug PRR, this adaptation for drug combinations
evaluates whether a combination represents a synergistic risk.
Specifically, as discussed by (Wang et al. 2020), a signal is often
defined by comparing the lower bound of the confidence interval of the
combination's risk to the maximum risk observed for the individual drug
components.
\item[\(\Omega\) Shrinkage]
A Bayesian measure specifically developed for drug-drug-event
combinations. As proposed by (Norén et al. 2008), it utilizes a
``shrinkage'' approach to reduce false positive signals in cases with
limited data by pulling the observed association toward the value
expected under independence.
\item[Concomitant Signal Score (CSS)]
Introduced as an improved detection criterion, this metric aims to
enhance the detection of DDI signals by building upon the PRR framework,
as detailed by (Noguchi et al. 2020).
\item[Hypergeometric Disproportionality Score]
While the methodology can accommodate various scores, we propose the use
of a score derived from the Fisher exact test. As highlighted by (Ahmed
et al. 2010), this approach has the advantage of being non-asymptotic.
Consider a dataset of \(N\) patients, among which \(K\) experience the
adverse event AE. Let \(n_C\) be the number of people taking a cocktail
\(C\) and \(x\) the number of patients taking \(C\) and experiencing AE.
We define the risk to be:
\end{description}

\[H(C) = -\log(\mathbb{P}(X \ge x))\]

where \(X \sim \mathcal{H}(n_C, K, N)\). This measure effectively
functions as the log p-value under the null hypothesis that the number
of people with AE in a uniform sample of \(n_C\) out of \(N\) people
follows the hypergeometric distribution. It has the distinct advantage
of accounting for the total number of patients exposed to the cocktail;
for instance, \(H(C)\) will assign a higher risk to a cocktail taken by
100 patients with 50 AE cases than to a cocktail taken by 10 patients
with 5 AE cases, whereas simple ratios like the RR might treat them
identically. Such hypergeometric measures are well-established in other
high-dimensional fields like bioinformatics for functional enrichment
analysis (Grossmann et al. 2007).

\subsection{Distinguishing Combined Risk from True
Interaction}\label{sec-firth}

While the \(H(C)\) score effectively identifies cocktails associated
with a significant increase in the AE frequency, it does not inherently
distinguish a synergistic pharmacological interaction from situations
corresponding to the independent addition of individual effects or an
effect driven by a subset of the cocktail (e.g., ``innocent''
medications co-prescribed with a high-risk drug).

To address this, we propose a post-treatment of detected signals using
Firth's penalized likelihood logistic regression (Firth 1993). This
method is particularly suited for spontaneous reporting systems as it
reduces bias in maximum likelihood estimates and provides a solution to
the problem of separation, which often occurs with rare adverse events.
Using the \emph{logistf} R package (Heinze, Ploner, and Dunkler 2023),
we estimate a model for a given cocktail \(C\) (e.g., \(d_1, d_2\)) as
follows:

\[\text{logit}(P(AE=1)) = \beta_0 + \beta_1 \mathbb{1}_{d_1} + \beta_2 \mathbb{1}_{d_2} + \beta_{3} \mathbb{1}_{d_1 \cap d_2}\]

An interaction is confirmed if the interaction coefficient \(\beta_{3}\)
is positive and statistically significant. This indicates that the risk
associated with the co-prescription exceeds the additive effect of the
individual components on the logit scale. While the example above
illustrates the case of a size-two cocktail for simplicity, this
post-treatment generalizes naturally to cocktails of any size \(k\) by
including all \(2^k - 1\) interaction terms in the model, or by testing
nested models comparing the full cocktail to its sub-cocktails. Users
may implement the post-treatment of their choice.

\subsection{Disproportionality
Identification}\label{disproportionality-identification}

\subsubsection{High-risk Drug Cocktails
Identification}\label{sec-genetic}

The number of possible cocktails is \(2^L\), where \(L\) denotes the
number of nodes in the ATC tree, and even for a given cocktail size
\(k\), the number of possibilities is still \(\binom{L}{k}\). It is
therefore not possible to compute \(H(C)\) for each possible cocktail
\(C\) included in the dataset. Instead, to explore the space of
cocktails and locate those at high-risk of AE, we use a genetic
algorithm (Pétrowski and Ben-Hamida 2017). It simulates the evolution of
a population of cocktails according to the principle of natural
selection, in order to search for the most performing individuals with
respect to an evaluation criterion based on \(H\).

The steps are the following, the algorithm repeating steps 2 and 3 until
a user-defined number of iterations is reached.

\begin{figure}

\centering{

\pandocbounded{\includegraphics[keepaspectratio]{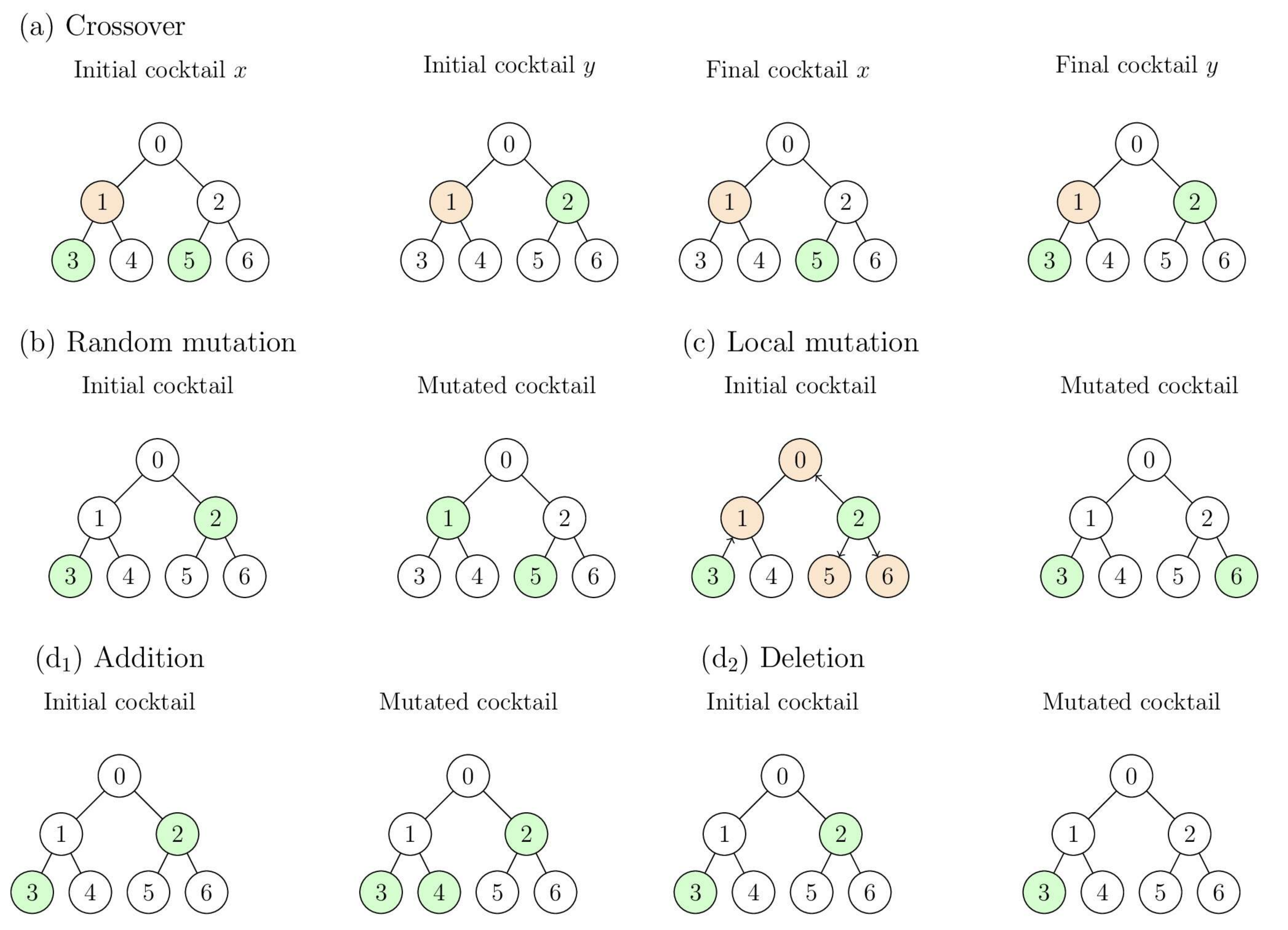}}

}

\caption{\label{fig-mut}Cocktail modifications used for the genetic
algorithm (a, c and d) and the MCMC algorithm (b and c). Green nodes are
part of considered cocktails. In Crossover, the orange node represents
the selected internal node whose subtrees are being swapped. In local
mutations, orange nodes represent legal moves}

\end{figure}

\begin{description}
\item[Initialization.]
The genetic algorithm's population consists of \(m\) cocktails. These
cocktails are randomly initialized and can vary in size.
\item[Evaluation and selection.]
At iteration \(n\), the current population \(P_n\) undergoes an
evaluation and selection phase. The evaluation computes the score
\(H(C)\) for each cocktail \(C\) in the population. A new population
\(Q_{n+1}\) of the same size is drawn by sampling \(m\) times a pair of
cocktails in the original population and copying the one with highest
score. Note that this step performs selection as the expectation of the
number of copies in \(Q_{n+1}\) of a cocktail from \(P_n\) is an
increasing function of its score. A penalization is however applied to
avoid a uniformization of the population, as further explained in
Section~\ref{sec-diversity}.
\item[Stochastic modification.]
Mimicking the genetic drift in nature, stochastic modifications occur in
the population in order to explore the large space of cocktails. First,
a crossover operation allows two cocktails to exchange information.
Here, the crossover involves exchanging sub-trees between two cocktails
as follows:

\begin{itemize}
\tightlist
\item
  an internal node \(v\) of the ATC tree is randomly selected;
\end{itemize}

\begin{itemize}
\tightlist
\item
  the nodes of the subtree rooted at \(v\) are exchanged between the two
  sequences.
\end{itemize}

After performing the crossover, a mutation is applied to the resulting
individuals, chosen from two types. The first type is a local mutation
which changes a randomly selected node of the cocktail to one of its
free neighboring nodes. This mutation is further explained in
Section~\ref{sec-MCMC}. The second type is an addition/deletion mutation
which operates as follows, with \(k\) being the cocktail length and
\(\alpha\) a chosen hyperparameter:

\begin{itemize}
\tightlist
\item
  with probability \(\min(1,\frac{\alpha}{k})\), a node uniformly drawn
  from \(\Delta\) is added to the sequence;
\end{itemize}

\begin{itemize}
\tightlist
\item
  with probability \(1 - \min(1,\frac{\alpha}{k})\), a uniformly drawn
  node from the cocktail is removed.
\end{itemize}

An example of crossover and mutations can be seen on
Figure~\ref{fig-mut} (a), (c) and (d).

Applying those modifications to \(Q_{n+1}\) yields a population
\(P_{n+1}\) which is used to loop at step \(n + 1\).
\end{description}

\subsubsection{Distance Between Drug Cocktails and Cocktail
Penalization}\label{sec-diversity}

If at some point of the algorithm, an individual of \(P_n\) has a high
score compared to other individuals, the genetic algorithm may converge
to a population consisting entirely of that cocktail. To avoid this
uniformization phenomenon, similar cocktails are penalized in the
evaluation phase as follows,

\[H_{pen}(C) = \frac{H(C)}{\sum_{C_i \in \mathcal{C}} \text{Sim}(C,C_i) }\]

The computation of the similarity \(\text{Sim}(C,C')\) is based on a
distance inspired by the Levenshtein distance (Levenshtein et al. 1966).
However, unlike the traditional Levenshtein distance, sequences are
treated as unordered sets. For two drug cocktails, \(C_1\) and \(C_2\),
of sizes \(n_1\) and \(n_2\), the distance \(d(C_1 , C_2)\) is defined
as the minimal cost required to transform \(C_1\) into \(C_2\) using
three elementary operations.

\begin{itemize}
\tightlist
\item
  \(\text{Ins}_a(C)\) consists of adding \(a\) to the cocktail \(C\).
\item
  \(\text{Del}_a(C)\) consists of deleting \(a\) from the cocktail
  \(C\).
\item
  \(\text{Sub}_{a,b}(C)\) consists of substituting \(a \in C\) by \(b\).
\end{itemize}

An associated cost is defined for each operation.

\begin{description}
\item[Substitution.]
The cost associated with the substitution operation is chosen to be
consistent with the conceptual similarity of cocktails. If \(a\) is a
drug belonging to \(C\), the cost should increase as the drug \(b\)
diverges further from drug \(a\). For example, if \(a\) is a drug, and
\(b\) a drug family that contains \(a\), the cost should be moderate.
Conversely, if \(b\) is a drug family not containing \(a\) the cost
should be higher. This distance is thus defined by the maximal distance
between \(a\) and \(b\) to their Lowest Common Ancestor.
\item[Insertion, Deletion.]
The deletion and insertion cost are chosen as \(\frac{depth(T)}{2}\).
This choice implies that a substitution always costs less than a
deletion followed by an insertion. The latter are used only when the two
cocktails do not have the same length.
\end{description}

A transformation \(f\) from \(C_1\) to \(C_2\) is a composition of
elementary operations that go from \(C_1\) to \(C_2\). The associated
cost \(\text{cost}(f)\) is defined as the sum of the cost of the
operations used in \(f\). Finally,
\(d(C_1,C_2) = \min\limits_{f} \text{cost}(f : f(C_1) = C_2 )\).

Finally, the maximum distance between \(C_1\) and \(C_2\) being
\((n_1 + n_2) \frac{depth(T)}{2}\), we define the similarity as
\[ Sim(C_1,C_2) = 1 - \frac{2d(C_1,C_2)}{(n_1 + n_2)depth(T)}\]

The computation of the similarity is achievable in
\(\mathcal{O}(n_1 \times n_2 \times   \text{depth}(T) + |\Delta|)\)
operations in the worst case. The algorithm to compute the similarity is
detailed in Section~\ref{sec-appA} (see Algorithm~\ref{algo-distance}).

\subsubsection{Output Clustering}\label{sec-clust}

Despite the diversity mechanisms integrated into the genetic algorithm,
not all drug cocktails retrieved within the genetic algorithm population
are unique. It is moreover common to encounter solutions that are merely
variations of others, differing only by transformations such as changing
a node in the ATC tree to its parent or child. To streamline analysis
and enhance the efficiency, a post-treatment clustering of similar drug
cocktails is implemented. This method allows focusing on the most risky
cocktails within each cluster or pharmaceutically interpreting clusters
rather than individual cocktails.

To do so, drug cocktails are embedded into a two-dimensional space using
the UMAP algorithm (McInnes, Healy, and Melville 2018), which aims to
preserve similarity in the latent space. This representation enables the
effective use of conventional machine learning clustering algorithms in
\(\mathbb{R}^2\). Specifically, the DBSCAN algorithm (Ester et al. 1996)
is applied to identify clusters of similar drug cocktails with an
intuitive way of choosing hyperparameters.

\subsection{Approximate P-Value Assignment for Drug
Cocktails}\label{sec-MCMC}

Once a list of cocktails with high score has been established by the
genetic algorithm, an important step of the analysis is to assign them
p-values to decide if they are significant.

Consider a cocktail \(C\) of size \(k\) in the list and its score
\(S_{obs}\). Denote by \(H_0\) the null hypothesis according to which a
cocktail does not favor the AE, and by \(N_0\) the number of cocktails
of size \(k\) for which \(H_0\) is the truth. Similarly, \(H_1\)
represents the alternative hypothesis of a cocktail favoring the AE and
\(N_1\) is the number of such cocktails of size \(k\). The p-value
corresponding to \(S_{obs}\) is then \(\mathbb{P}_{H_0}(S>S_{obs})\),
but cannot be estimated without a modelling hypothesis under \(H_0\).

However, if \(\mathbb{P}_M (S)\) refers to the marginal distribution,
that is the score distribution for the cocktails both in \(H_0\) and
\(H_1\),

\[\mathbb{P}_M(S>S_{obs})  = \mathbb{P}_{H_0}(S>S_{obs}) \frac{N_0}{N_0+N_1} + \mathbb{P}_{H_1}(S>S_{obs}) \frac{N_1}{N_0+N_1}\]
so that

\[\frac{N_0+N_1}{N_0}\mathbb{P}_M(S>S_{obs}) - \frac{N_1}{N_0} \mathbb{P}_{H_1}(S>S_{obs}) \leq  \mathbb{P}_{H_0}(S>S_{obs}) \leq \frac{N_0+N_1}{N_0}\mathbb{P}_M(S>S_{obs})\]
Under the reasonable assumption \(N_1 << N_0\), the probability
\(\mathbb{P}_M(S>S_{obs})\) on the marginal distribution can thus be
used as an approximation of the p-value. Note that the upper bound
yields that, under the weak hypothesis that less than \(10\%\) of the
cocktails of size \(k\) favor the AE, the real p-value is at most equal
to \(1.11\) times its approximation. Moreover, the approximated p-value
can be estimated by computing the score of cocktails sampled in the
general cocktail population. The proposed method therefore focuses on a
sampling scheme in the general population and uses the approximated
p-value to declare significance.

A naive sampling of cocktails considers almost only cocktails taken by
no patient in the dataset. A Metropolis-Hastings MCMC algorithm is thus
considered, as it can be used by conditioning on the fact that all
visited cocktails are present in the dataset (Au and Beck 2001).

To employ such an algorithm, it is necessary to define a state space
\(\mathcal{C} = \{C_{1}, \dots, C_{p}\}\), a computable target measure
\(f(C_{i})\), and conditional laws \(q(.|C_{i})\) under which simulation
is possible and new states can be proposed.

\begin{description}
\item[State set.]
The state set is made of all cocktails of \(k\) drugs for a fixed \(k\).
\item[Proposal law.]
The proposal law is defined as a mixture of two mutation laws of the
current cocktail. They operate as follows:

\begin{itemize}
\tightlist
\item
  Random mutation consists of a completely random movement in the
  cocktail space.
\end{itemize}

\begin{itemize}
\tightlist
\item
  Local mutation involves a local movement relative to the structure of
  the drug tree. Here, a node \(x_p\) of the state \(C_i\) is changed to
  one of its free neighboring nodes.
\end{itemize}
\end{description}

At each iteration, the random and local mutations have probability
\(p_R\) and \(1-p_R\), where \(p_R\) is a hyperparameter.

Figure~\ref{fig-mut} (b, c) presents examples of a random and a local
mutation.

\begin{description}
\item[State evaluation.]
The evaluation of a drug cocktail is based on the score \(H(C)\). The
chosen target measure is then:
\[f_{T}(C_i) = \frac{1}{Z(T)} \times {e^{\frac{H(C_i)}{T}}}\]

where \(Z(T) = \sum_{C} e^{\frac{H(C)}{T}}\). \(T\) is a parameter known
as temperature, which modulates space exploration by more readily
accepting cocktails with moderate scores (high \(T\)) or, conversely, by
strongly favoring combinations of drugs with high scores (low \(T\)).

The acceptance probability of cocktail \(C_{i+1}\) from cocktail
\(C_{i}\) is given by:
\end{description}

\[\min\left(1,\frac{f_T(C_{i+1})}{f_T(C_{i})} \times \frac{q(C_{i}|C_{i+1})}{q(C_{i+1}|C_{i})}\right)\]

The theory related to the Metropolis-Hastings algorithm (Robert and
Casella 2004) ensures that the empirical distribution of \(f_T(C_i)\)
for the constructed cocktail chain converges to the distribution of
\(f_T(C)\). A very long realization of such a walk, therefore, allows
for the approximation of the distribution that can be used to determine
an empirical p-value for the score of a cocktail of interest. This
enables the possibility to say whether or not a high-risk is truly
significant (e.g.~among the top 5\% of scores). It defines what a
high-risk is in our case.

\subsection{Datasets}\label{sec-datasets}

\subsubsection{Simulated Data}\label{sec-simu_data}

Multiple datasets were simulated to evaluate the method's performance
against known outcomes. The datasets, designed to challenge the
algorithm, simulate various patient scenarios. Each patient record
includes prescribed medications and the corresponding occurrence of an
adverse event \(AE\).

The first dataset (\(D_1\)) is composed of 200,000 patients, and has the
following characteristics:

\begin{itemize}
\tightlist
\item
  \(1\%\) of the patients take a size-3 drug cocktail \(C_1\) and have a
  \(\frac{1}{100}\) chance of having \(AE\).
\item
  \(1\%\) take a size-3 drug cocktail \(C_2\) and have a
  \(\frac{1}{200}\) chance of having \(AE\).
\item
  \(1\%\) take a size-2 drug cocktail \(C_3\) and have a
  \(\frac{1}{100}\) chance of having \(AE\).
\item
  \(1\%\) take a size-2 drug cocktail \(C_4\) and have a
  \(\frac{1}{200}\) chance of having \(AE\).
\end{itemize}

A small percentage of the dataset (\(1.5 \%\) per combination) are
combinations of 2 out of the 3 drugs from \(C_1\) and \(C_2\), but with
no risk of \(AE\). This helps to mitigate the false identification of
sub-cocktails of \(C_1\) and \(C_2\) as high-risk cocktails because
those who take two drugs of these cocktails will almost surely take the
remaining drug of the cocktail.

The remaining \(87\%\) of the dataset consists of patients assigned
random cocktails drawn uniformly. The size \(s\) of each cocktail is
drawn according to a Poisson distribution with \(\lambda = 4\) (mean
size of drugs cocktails taken by patients in the dataset). For each
cocktail, \(s\) nodes of the ATC tree are selected uniformly, with each
combination assigned an adverse event with probability
\(\frac{1}{15000}\).

All simulations use the complete ATC classification tree comprising 6809
nodes across five hierarchical levels at the time of this study. The
cocktails \(C_1\) through \(C_4\) inducing an elevated AE probability
were chosen as nodes at level 4 of the ATC hierarchy (i.e.,
pharmacological subgroups, one level above the leaf nodes representing
individual active substances). This choice is deliberate: in the
simulated data, patients are assigned specific drugs at level 5 (leaves)
as would occur in practice, but any patient taking a drug that descends
from one of the \(C_i\) nodes inherits the elevated AE probability
associated with that family. This design tests the method's ability to
detect family-level signals --- an important feature, since the
underlying pharmacological mechanism may apply to an entire drug class
rather than a single substance. The \(C_1\) through \(C_4\) nodes were
selected to be non-overlapping (i.e., belonging to distinct subtrees)
and to have a sufficient number of descendant leaves to generate
realistic prescription patterns.

Three additional datasets were similarly constructed, varying the sizes
of the cocktails inducing AE. Dataset \(D_1\) contains cocktails of
sizes two and three as described above. Dataset \(D_2\) contains only
size-two cocktails, Dataset \(D_3\) only size-three cocktails, and
Dataset \(D_4\) contains size-two, three, and four cocktails.

\subsubsection{FAERS Data}\label{sec-FAERS}

As a proof-of-concept on real data, the method was also assessed using
the FAERS dataset, which consists of ICSRs submitted by healthcare
professionals, consumers, and manufacturers. These reports include
details on patient drug intake and the side effects experienced. The
methods were deployed on FAERS data from the second quarter of \(2013\)
to the second quarter of \(2015\), the restriction to a two-year window
being motivated by computational time constraints.

Significant refinement of the FAERS data was required. Duplicate reports
were removed, retaining only the report with the most recent ID.
Subsequently, a link was established between the prescribed drugs and
the ATC codes of each active ingredient. This process involved matching
drug names to their respective active ingredients and converting these
ingredients to their corresponding ATC codes. The DiAna dictionary
(Fusaroli et al. 2024) facilitated the standardization of FAERS drug
names to ATC codes. It is important to note that the DiAna dictionary
handles combination products by splitting them into their constituent
active substances rather than assigning a single ATC combination code.
Reports with unmatchable drug names in the DiAna dictionary were
excluded, reducing the dataset from 2,043,231 to 1,612,931 patients.

For this study, we focused on myopathy as the selected adverse event
outcome. It is a clinically concerning condition with a sufficient
number of reported cases in the dataset (\(536\) cases). To validate our
results, we compared the identified drug-myopathy associations with
known drugs already established to cause myopathy (Miernik, Matusiewicz,
and Olesińska 2024; Hall, Finnoff, and Smith 2011; Valiyil and
Christopher-Stine 2010). Code for data refinement is available
\href{https://github.com/JulesBa-Git/FAERS-xml-parser}{on GitHub}.

\section{Results}\label{results}

\subsection{Score Comparison}\label{sec-score_comp}

\begin{figure}

\centering{

\begin{figure}[H]

\begin{minipage}{0.50\linewidth}
\pandocbounded{\includegraphics[keepaspectratio]{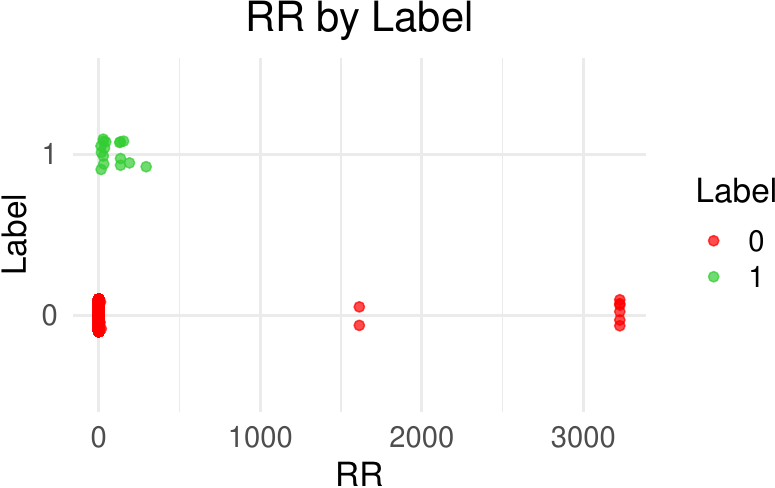}}\end{minipage}%
\begin{minipage}{0.50\linewidth}
\pandocbounded{\includegraphics[keepaspectratio]{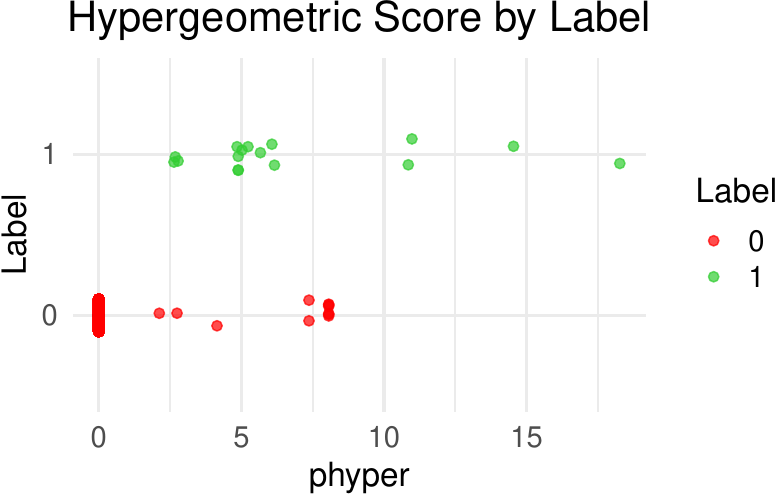}}\end{minipage}%
\newline
\begin{minipage}{0.50\linewidth}
\pandocbounded{\includegraphics[keepaspectratio]{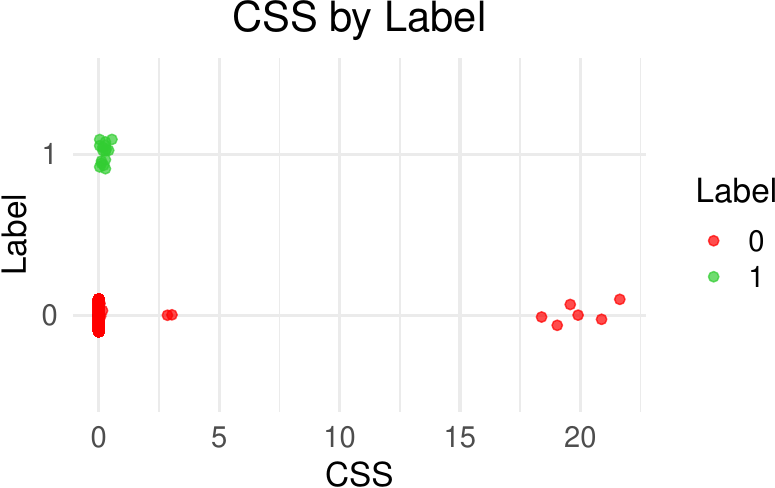}}\end{minipage}%
\begin{minipage}{0.50\linewidth}
\pandocbounded{\includegraphics[keepaspectratio]{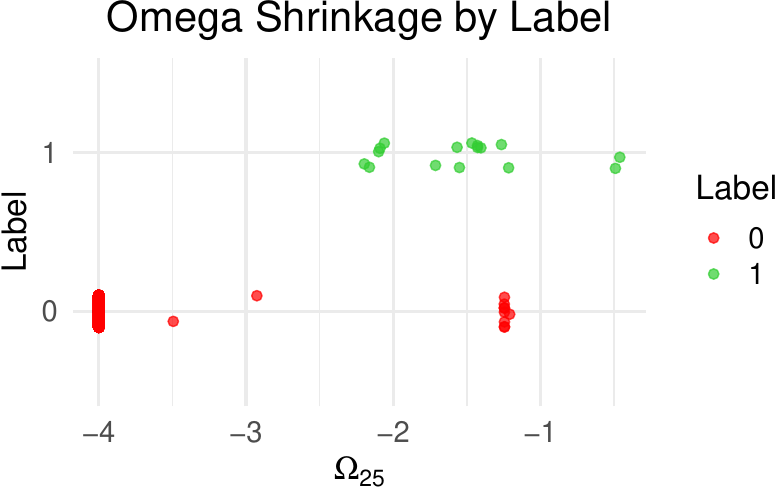}}\end{minipage}%
\newline
\begin{minipage}{0.10\linewidth}
~\end{minipage}%
\begin{minipage}{0.80\linewidth}
\pandocbounded{\includegraphics[keepaspectratio]{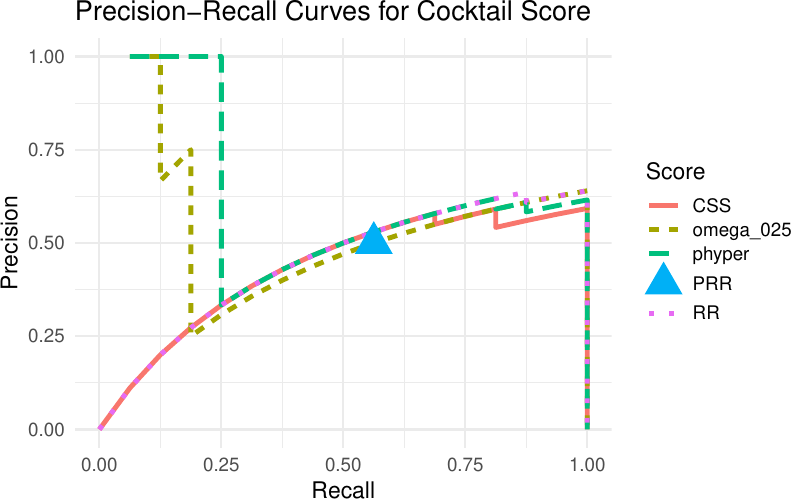}}\end{minipage}%
\begin{minipage}{0.10\linewidth}
~\end{minipage}%

\end{figure}%

}

\caption{\label{fig-comparison_score}Comparison of scores for cocktails
of size-two. Each dot denotes a cocktail, and its risk computed on
Dataset \(D_2\) (Section~\ref{sec-simu_data}) is shown on the x-axis.
Red dots represent cocktails that do not induce adverse event (negative
controls), green ones represent cocktails inducing adverse event
(positive controls). The bottom-right corner shows the Precision-Recall
curves for each score. The perfect classification corresponds to the
upper right corner. The areas under the curves allow comparison of
different methods. RR, PRR, Omega shrinkage, phyper and the CSS are
defined in Section~\ref{sec-risk-definition}.}

\end{figure}

To support the use of the hypergeometric score \(H(C)\) introduced in
Section~\ref{sec-risk}, we compared it to other risk scores on Dataset
\(D_2\), which comprises pairwise drug combinations. A set of cocktails
was generated under the alternative hypothesis \(H_1\), meaning they
correspond to the drug combinations specifically designed to increase
the probability of AE in Dataset \(D_2\) (see
Section~\ref{sec-simu_data}). A second set of cocktails was sampled
under the null hypothesis \(H_0\), representing a baseline risk without
specific interaction. For each cocktail in both sets, the risk scores
were computed using the patient data in Dataset \(D_2\).
Figure~\ref{fig-comparison_score} illustrates the performance of the
risk scoring methods for detecting high-risk drug combinations
introduced in Section~\ref{sec-risk-definition}. Each subplot displays
the score values for cocktails representing true solutions, sampled
under \(H_1\) (green) and cocktails not representing true solutions,
sampled under \(H_0\) (red). The bottom-right panel presents the
Precision-Recall (PR) curve, comparing the detection power of the scores
in identifying high-risk cocktails. PRR is a special score because its
value is either one or zero, which allows the computation of only one
value of precision and recall. \(PRR\) is therefore represented by a
single point rather than a PR curve.

PR curves of the different scoring methods are mainly comparable,
confirming the conclusions of (Candore et al. 2015). However, as
illustrated by the jitter plots, the hypergeometric score and the
\(\Omega\) Shrinkage measure better rank and isolate true solutions from
other cocktails. It must be emphasized that the \(\Omega\) Shrinkage
measure is difficult to compare using a threshold, as the original
article (Norén et al. 2008) suggests signaling a cocktail when the score
exceeds zero. However, in the simulations, the computed score never
exceeded zero, as depicted by the x-axis of its jitter plot. Despite
these considerations, these two methods are the least biased toward
cocktails taken by only a few patients.

The good behavior of the hypergeometric score and its easy
generalization to cocktails of any size justify the use of that scoring
function for the exploration of the cocktail space.

\subsection{\texorpdfstring{Application to Dataset
\(D_1\)}{Application to Dataset D\_1}}\label{sec-appli_d1}

\subsubsection{Estimation of Risk
Distribution}\label{estimation-of-risk-distribution}

\begin{figure}

\centering{

\begin{figure}[H]

\begin{minipage}{0.50\linewidth}
\pandocbounded{\includegraphics[keepaspectratio]{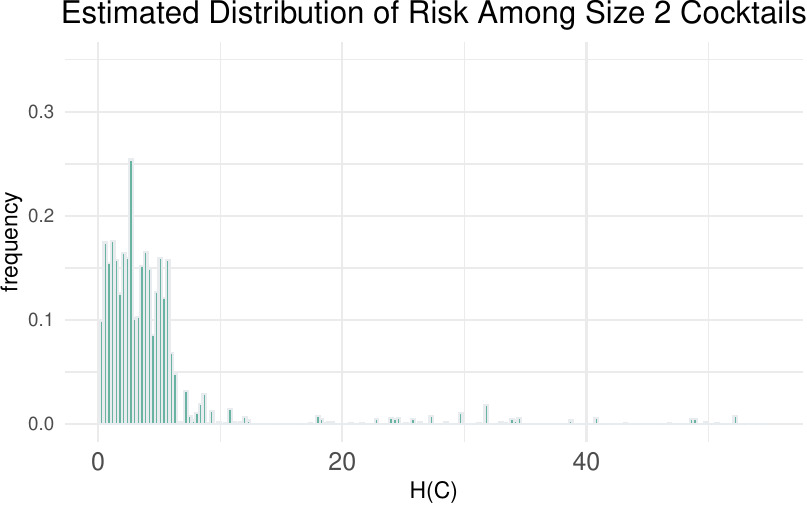}}\end{minipage}%
\begin{minipage}{0.50\linewidth}
\pandocbounded{\includegraphics[keepaspectratio]{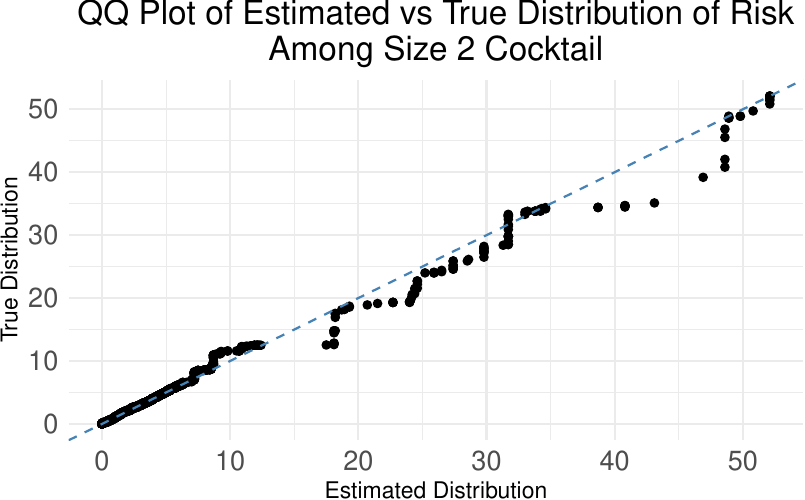}}\end{minipage}%
\newline
\begin{minipage}{0.50\linewidth}
\pandocbounded{\includegraphics[keepaspectratio]{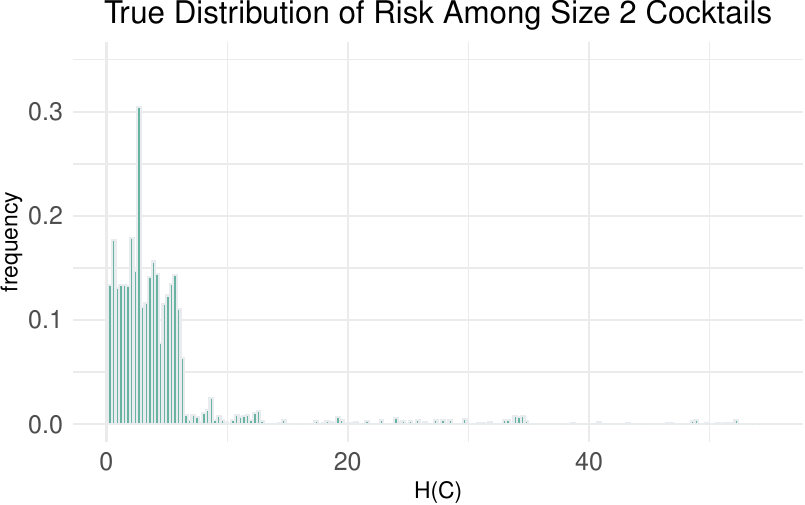}}\end{minipage}%
\begin{minipage}{0.50\linewidth}
\pandocbounded{\includegraphics[keepaspectratio]{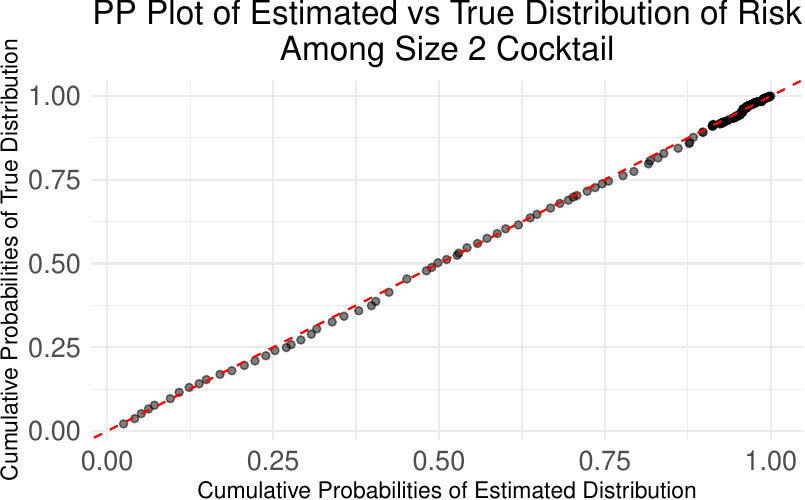}}\end{minipage}%

\end{figure}%

}

\caption{\label{fig-distri_simu}Comparison of estimated and true risk
distributions for size-two drug cocktails on Dataset \(D_1\). Left
panels show comparison of risk distribution among size-two cocktails,
right panels allow comparison of probabilities and quantiles of both
distributions.}

\end{figure}

Risk distribution was estimated for size-two drug cocktails on Dataset
\(D_1\) (Section~\ref{sec-simu_data}). Estimation of risk distribution
for higher cocktail sizes is possible but it is nearly impossible to
compare it to the true distribution as it is computationally prohibitive
to obtain. The distribution estimated by the MCMC algorithm is compared
to the true risk distributions in Figure~\ref{fig-distri_simu}.

The left panels display the distributions of risk scores for both the
estimated (top-left) and true (bottom-left) risk values. Both
distributions share a similar shape, with the majority of risk scores
concentrated at low values, as \(95\%\) of the scores fall below 11.
However, some differences can be observed, particularly in the tail of
the distribution, where cocktails of higher risks are under-represented
in the estimated distribution.

The right panels of Figure~\ref{fig-distri_simu} present a
Probability-Probability (PP) plot (top) and a Quantile-Quantile (QQ)
plot (bottom), comparing the quantiles and probabilities of the
estimated and true risk distributions. While the right panel of the
figure demonstrates good agreement at lower risks, where most of the
data lie, deviations at higher risk values suggest that the estimated
distribution slightly underrepresents the risk for more extreme values.

Results indicate that the method performs well in estimating risk scores
for the majority of cocktails, capturing the overall risk distribution
with reasonable accuracy. The slight underestimations which are present
for high-risk cocktails are not a problem since the interest of the
method is to assign p-values. P-values are still reliable as shown in
the PP-plot, Figure~\ref{fig-distri_simu}. Consequently, we can assign
robust empirical p-values to any size-two cocktail based on its risk.
Furthermore, the calibration under the null hypothesis for the p-values
attributed by the algorithm is assessed in Section~\ref{sec-appB}.

\subsubsection{Genetic Algorithm Output and
Clustering}\label{genetic-algorithm-output-and-clustering}

The genetic algorithm was applied to Dataset \(D_1\) to identify
high-risk drug cocktails. Multiple runs of the algorithm were conducted
using different hyperparameter sets (population size, number of
generations, parameter \(\alpha\)) to ensure the visit of different
regions of the space of possible cocktails. The results were
subsequently concatenated to create a comprehensive list of high-risk
cocktails.

\begin{figure}

\centering{

\includegraphics[width=0.9\linewidth,height=\textheight,keepaspectratio]{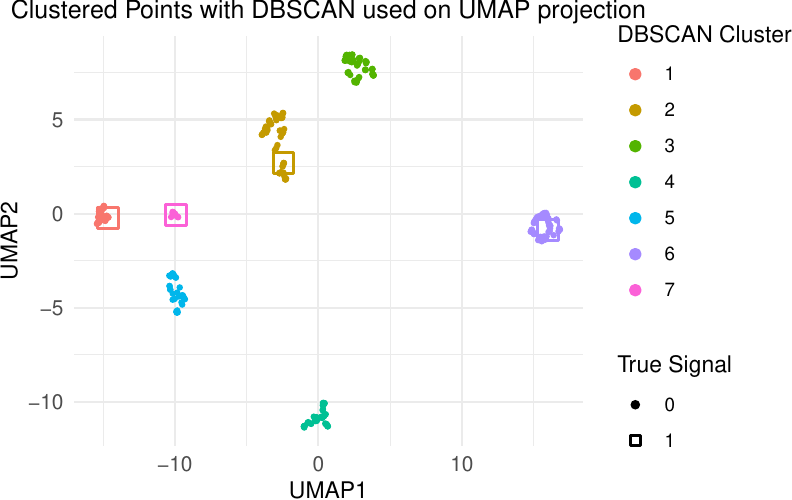}

}

\caption{\label{fig-cluster_simu}Clustering of high-risk drug cocktails
identified by the genetic algorithm on the simulated dataset. Each dot
represents a drug cocktail, true cocktail solutions are the centers of
the squares}

\end{figure}

The genetic algorithm successfully identified nearly all high-risk
size-two and size-three drug cocktails in Dataset \(D_1\)
(Figure~\ref{fig-cluster_simu}). For size-two cocktails, the algorithm
consistently found the exact high-risk combinations. However, for
size-three cocktails, the algorithm sometimes identified cocktails that
were very close to the true high-risk combinations, missing only one
drug from the correct cocktail in a few cases (oftentimes, choosing
parent nodes instead of the actual nodes).

To streamline the analysis of the large set of results, significant
cocktails were filtered using the empirical p-value by setting a
threshold of \(5\%\). Clustering techniques were then applied to group
similar cocktails together. As discussed in Section~\ref{sec-clust}, the
UMAP algorithm has been used for dimensionality reduction, followed by
the DBSCAN clustering method. This post-processing step allows reducing
redundancy by grouping cocktails that differ only slightly, such as by
substituting a drug for another within the same pharmacological family.
Such cocktails would have similar medical interpretations.

Figure~\ref{fig-cluster_simu} illustrates the results of the clustering
process after mapping the selected cocktails in the plane. Each point
represents a drug cocktail, and the \(7\) automatically determined
clusters are represented by the colors.

The first conclusion is that the clusters are well separated in terms of
the editing distance defined in Section~\ref{sec-diversity}. The genetic
algorithm thus identified seven regions of the cocktails space of high
score that are different in terms of interpretation as their cocktails
are far apart across classes. On the other hand, cocktails in the same
clusters are close, indicating a possible pharmacological interpretation
of the cluster.

The second conclusion is that the algorithm effectively found the true
solutions or at least very similar solutions. Indeed, the embedding of
the true solutions yields the four squares, that clearly belong to four
different clusters among the seven. The method therefore identifies and
separates regions of the cocktail space containing the true solutions.
That simulation thus enhances the confidence that investigating the
high-score cocktails returned by the methods may allow detecting
relevant phenomena.

\subsection{Application to the FAERS Spontaneous Reporting
Data}\label{sec-appli-FAERS}

\subsubsection{Estimation of Risk
Distribution}\label{estimation-of-risk-distribution-1}

\begin{figure}

\centering{

\begin{figure}[H]

\begin{minipage}{0.50\linewidth}
\pandocbounded{\includegraphics[keepaspectratio]{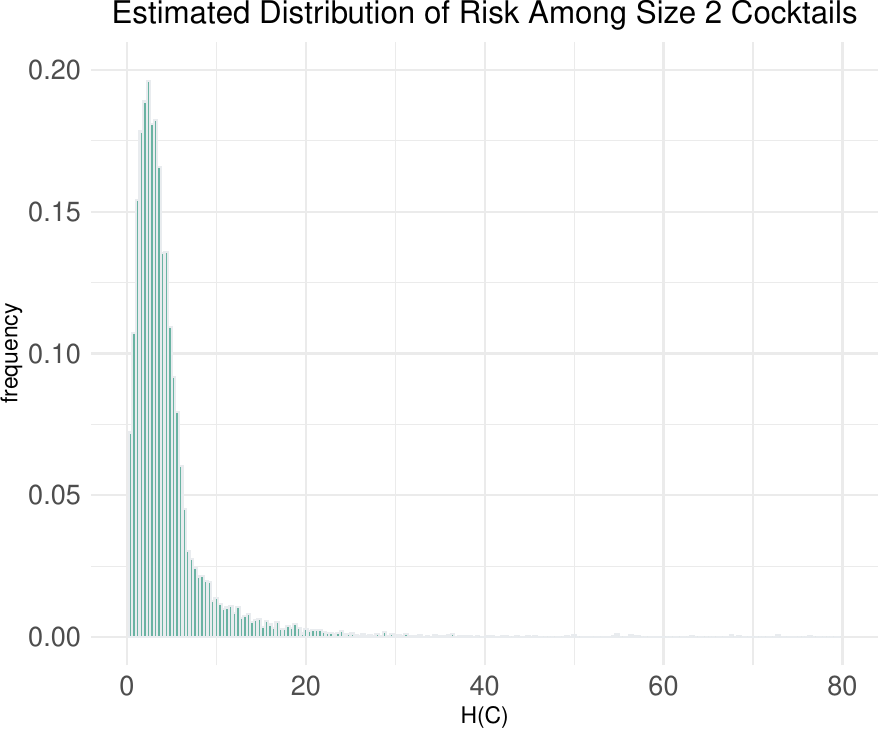}}\end{minipage}%
\begin{minipage}{0.50\linewidth}
\pandocbounded{\includegraphics[keepaspectratio]{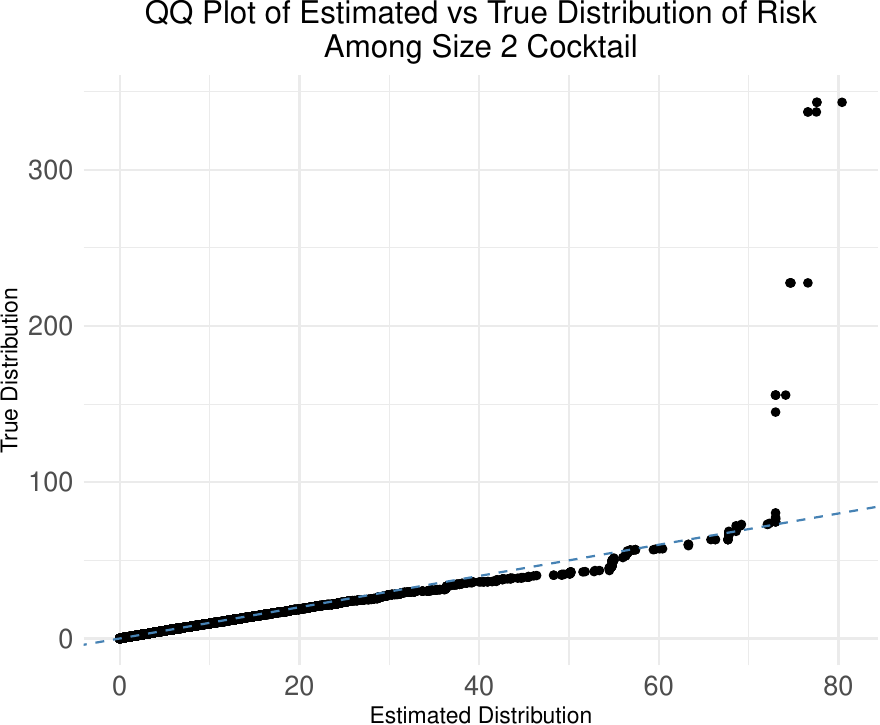}}\end{minipage}%
\newline
\begin{minipage}{0.50\linewidth}
\pandocbounded{\includegraphics[keepaspectratio]{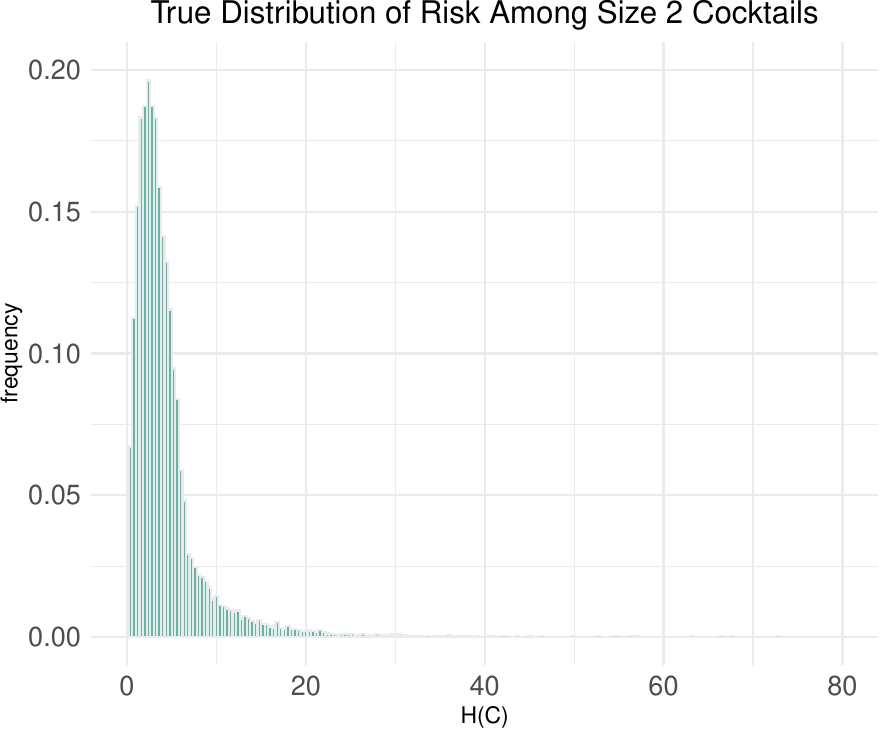}}\end{minipage}%
\begin{minipage}{0.50\linewidth}
\pandocbounded{\includegraphics[keepaspectratio]{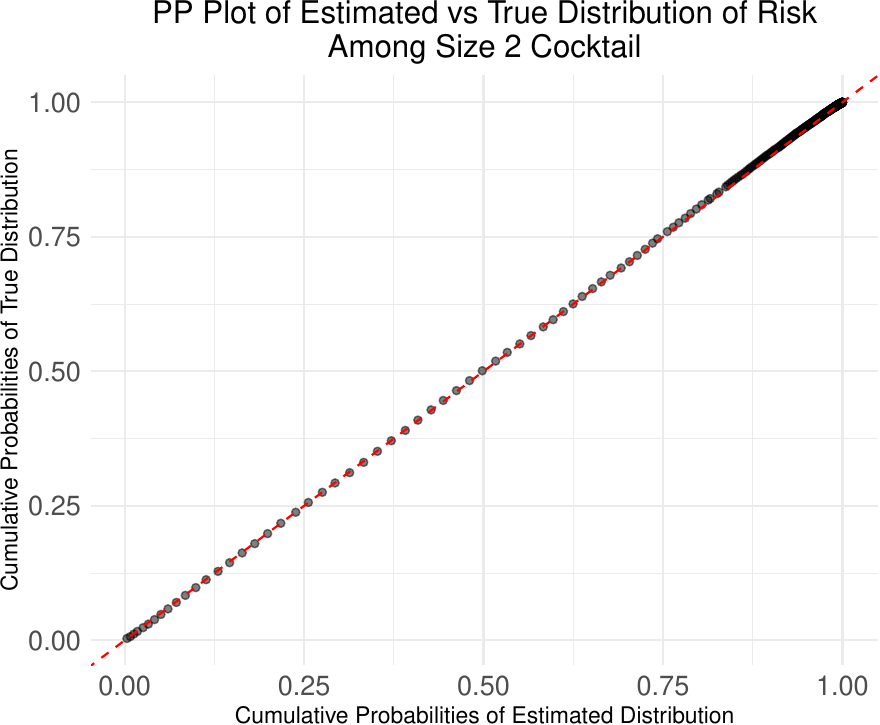}}\end{minipage}%

\end{figure}%

}

\caption{\label{fig-qqplot_FAERS}Comparison of estimated and true risk
distributions for size-two drug cocktails on the FAERS dataset presented
in Section~\ref{sec-datasets}. Left panels show comparison of risk
distribution among size-two cocktails, right panels allow comparison of
probabilities and quantiles of both distributions. Left distributions
are truncated to \(80\) for visualization purposes. A few cocktails have
a risk around \(300\) as shown on the Y-axis of the QQ plot.}

\end{figure}

The risk estimation method was applied to the FAERS spontaneous
reporting dataset presented in Section~\ref{sec-FAERS}.
Figure~\ref{fig-qqplot_FAERS} presents a comparison between the
estimated risk distribution and the true risk distribution for size-two
drug cocktails. The left panel of the figure shows the distributions of
risk scores, while the right panel presents the QQ plot and the PP plot,
comparing the quantiles and probabilities of both distributions.

The histogram reveals that the estimated distribution aligns well with
the true distribution for the majority of cocktails with lower risk
scores. However, deviations begin to emerge in the tail of the
distribution. Specifically, 15 of the riskiest cocktails in the true
distribution were not captured by the estimated distribution as we see
in the QQ plot. This explains the observed shift in the QQ plot at the
highest quantiles since higher risk cocktails have not been found by the
MCMC algorithm, highlighting the need for a complementary method like
the genetic algorithm in order to find the riskiest cocktails.

Despite this slight deviation, the empirical p-values remain robust for
both lower and higher-risk cocktails as shown by the PP-plot in
Figure~\ref{fig-qqplot_FAERS}.

\subsubsection{Genetic Algorithm Output and
Clustering}\label{genetic-algorithm-output-and-clustering-1}

The genetic algorithm was applied to the FAERS data focusing on the
myopathy AE by running \(180\) parallel executions of the genetic
algorithm on varying population sizes (from \(100\) to \(1000\)
cocktails per generation). The whole procedure ran in less than \(8\)
hours on a \(24\)-core server.

Cocktail sizes present in the merged final populations vary from \(1\)
to \(6\). The MCMC algorithm was run for each cocktail size in this
range to assign an empirical p-value to each solution. All solutions
having a p-value lower than \(0.05\) were kept. No multiple testing
correction was made in order to avoid false negatives, that is
disregarding interesting cocktails, even if this may inflate the number
of false positives. \(682\) drug cocktails compose the final list.
Applying Benjamini-Hochberg corrections to control the FDR at level
\(5\%\) and \(10\%\) respectively kept the first \(107\) and \(564\)
elements of the list.

The clustering procedure was applied to the \(682\) cocktails, leading
to \(15\) clusters. To validate the results, the first \(150\) solutions
were tagged with the families they correspond to, if any, without
knowledge of the clustering's result. To do so, some drugs or drug
families known to be linked to myopathy adverse events were considered,
that is hypolipemic drugs, Colchicine, corticosteroids, Cyclosporine,
Beta blocking agents, Fluoroquinolones, and anti-malaria drugs (Miernik,
Matusiewicz, and Olesińska 2024; Hall, Finnoff, and Smith 2011; Valiyil
and Christopher-Stine 2010).

The final result is a table of \(682\) rows, one per selected cocktail,
indicating its composition, the number of patients taking it, the number
of patients taking it and facing myopathy, hypergeometric and RR score.
It also contains the cluster it belongs to and the tagged families. The
entire table is available
\href{https://docs.google.com/spreadsheets/d/1TmbrExi26QNr2TmvltmEzyl_EQht3fIsrtsWrAqmRik/edit?gid=2108693123\#gid=2108693123}{on
this link}. Table~\ref{tbl-cluster_table} summarizes the cluster
assignments for the top \(150\) cocktails identified in the genetic
algorithm. Among these, eleven cocktails could not be assigned to any
drug or drug family listed in the table headers. Note that a cocktail
can be associated with more than one drug or drug family.

% Keep long column headings inside the table.
\begingroup
\small
\setlength{\tabcolsep}{3pt}
\begin{longtable}[]{@{}
  >{\centering\arraybackslash}p{(\linewidth - 14\tabcolsep) * \real{0.1250}}
  >{\centering\arraybackslash}p{(\linewidth - 14\tabcolsep) * \real{0.1250}}
  >{\centering\arraybackslash}p{(\linewidth - 14\tabcolsep) * \real{0.1250}}
  >{\centering\arraybackslash}p{(\linewidth - 14\tabcolsep) * \real{0.1250}}
  >{\centering\arraybackslash}p{(\linewidth - 14\tabcolsep) * \real{0.1250}}
  >{\centering\arraybackslash}p{(\linewidth - 14\tabcolsep) * \real{0.1250}}
  >{\centering\arraybackslash}p{(\linewidth - 14\tabcolsep) * \real{0.1250}}
  >{\centering\arraybackslash}p{(\linewidth - 14\tabcolsep) * \real{0.1250}}@{}}
\caption{Summary of clustering applied to the identified solutions. Each
row corresponds to a cluster. The \(150\) cocktails with the highest
risk were analyzed. Box \((i,j)\) in the table represents the number of
cocktails in cluster \(i\) that include the drug or drug family \(j\)
}\label{tbl-cluster_table}\tabularnewline
\toprule\noalign{}
\begin{minipage}[b]{\linewidth}\centering
Cluster
\end{minipage} & \begin{minipage}[b]{\linewidth}\centering
Hypo\-lipemic Drugs
\end{minipage} & \begin{minipage}[b]{\linewidth}\centering
Colchi\-cine
\end{minipage} & \begin{minipage}[b]{\linewidth}\centering
Steroids
\end{minipage} & \begin{minipage}[b]{\linewidth}\centering
Cyclo\-sporine
\end{minipage} & \begin{minipage}[b]{\linewidth}\centering
Beta blocking agents
\end{minipage} & \begin{minipage}[b]{\linewidth}\centering
Domperi\-done
\end{minipage} & \begin{minipage}[b]{\linewidth}\centering
Fluoro\-quino\-lones
\end{minipage} \\
\midrule\noalign{}
\endfirsthead
\toprule\noalign{}
\begin{minipage}[b]{\linewidth}\centering
Cluster
\end{minipage} & \begin{minipage}[b]{\linewidth}\centering
Hypo\-lipemic Drugs
\end{minipage} & \begin{minipage}[b]{\linewidth}\centering
Colchi\-cine
\end{minipage} & \begin{minipage}[b]{\linewidth}\centering
Steroids
\end{minipage} & \begin{minipage}[b]{\linewidth}\centering
Cyclo\-sporine
\end{minipage} & \begin{minipage}[b]{\linewidth}\centering
Beta blocking agents
\end{minipage} & \begin{minipage}[b]{\linewidth}\centering
Domperi\-done
\end{minipage} & \begin{minipage}[b]{\linewidth}\centering
Fluoro\-quino\-lones
\end{minipage} \\
\midrule\noalign{}
\endhead
\bottomrule\noalign{}
\endlastfoot
1 & \textbf{31} & 0 & 3 & 0 & 0 & 0 & 4 \\
2 & 0 & \textbf{11} & 0 & 4 & 0 & 0 & 0 \\
3 & 7 & 0 & 0 & 0 & \textbf{19} & 0 & 0 \\
4 & 0 & 0 & 0 & 0 & 0 & \textbf{50} & 0 \\
5 & 0 & 0 & 0 & 0 & 0 & \textbf{19} & 0 \\
6 & 0 & 0 & 0 & 0 & 0 & 0 & \textbf{4} \\
\end{longtable}
\endgroup

This table shows the method's capability to detect known ADRs from the
FAERS dataset and highlights the ability of the clustering method to
group cocktails with similar pharmacological interpretations.
Specifically, cluster \(1\) corresponds to cocktails containing
Hypolipemic Drugs, cluster \(2\) to Colchicine and Cyclosporine, cluster
\(3\) to Beta blocking agents and cluster \(6\) to Fluoroquinolones.
These clusters align with previously reported pharmacological
associations (Miernik, Matusiewicz, and Olesińska 2024). Similarly,
clusters \(4\) and \(5\) correspond to Domperidone-containing cocktails
which have been associated with adverse effects, including potential
cardiac complications, as reported by the British Medicines and
Healthcare products Regulatory Agency (Medicines and Healthcare products
Regulatory Agency 2014).

Interestingly, additional drug families reported in (Miernik,
Matusiewicz, and Olesińska 2024), such as anti-malarial drugs, were also
identified but at later ranks. For instance, anti-malarial drugs are
primarily grouped in cluster \(9\), as shown in the complete solution
table.

\subsubsection{Post-Treatment Interaction Analysis}\label{sec-resFirth}

The penalized logistic regression described in Section~\ref{sec-firth}
was applied to the strongest signals identified by the genetic algorithm
to assess whether the detected cocktails reflect true pharmacological
interactions. For the cyclosporine--colchicine combination, the
interaction coefficient remained positive and highly significant
(\(p = 1.02 \times 10^{-6}\)) after accounting for individual drug
effects, corroborating existing medical literature indicating a
potentiation of muscle toxicity (Ducloux et al. 1997).

Conversely, the analysis of the identified size-four cocktail
(metformin, prasugrel, bisoprolol, simvastatin) revealed that the risk
signal was almost entirely explained by a size-three sub-cocktail
(metformin, bisoprolol, simvastatin). In this instance, the addition of
prasugrel did not yield a significant additional interaction effect,
illustrating the importance of the post-treatment step in identifying
the core of a detected signal.

\section{Conclusion}\label{conclusion}

As co-medication becomes increasingly common, there is a growing need
for methods capable of detecting signals of harmful drug combinations
from the available large databases for further assessment. The proposed
method addresses this need by identifying signals and assigning them a
p-value using a hypergeometric disproportionality analysis measure.
Additionally, the method enables the identification of broader signals
within the ATC hierarchy by proposing ``cocktails'' of not only active
substances but also chemical, therapeutic, and anatomical families,
leveraging the hierarchical classification of active substances.

Application on synthetic datasets demonstrated that using the
hypergeometric score reduces the false positives from cocktails taken by
a small number of patients, enhancing the robustness of the measure. The
results on these datasets of our MCMC algorithm to estimate the
distributions of cocktail risks were encouraging, as the estimated
distributions closely aligned with the true distributions, indicating
reliable p-value assignment. Furthermore, the genetic algorithm
effectively identified the majority of the harmful cocktails, with a
high success rate, highlighting its efficiency in navigating the large
solution space.

Applying this method to previous FAERS data for the myopathy adverse
event yielded promising results. A literature review confirmed the
intersection between the identified signals and drugs known to have a
higher likelihood of causing myopathy, demonstrating the effectiveness
of the proposed methodology. Results also indicate that certain drug
combinations are more strongly associated with myopathy than individual
medications. Notably, the cyclosporine/colchicine combination exhibits a
higher hypergeometric score (73.1 vs.~63.4) and RR (879.4 vs.~57.1)
compared to colchicine alone, reinforcing the importance of analyzing
drug interactions in pharmacovigilance. This combination is known to be
more likely to induce myopathy (Ducloux et al. 1997).

Furthermore, our approach identified a size-four drug cocktail
(metformin, prasugrel, bisoprolol, simvastatin) associated with an
increased risk signal (hypergeometric score \(= 72.2\), RR =
\(3060.6\)). This combination was observed in nine patients, all of whom
experienced the adverse event. While this finding demonstrates the
feasibility of detecting higher-order drug interactions, we emphasize
that no clinical validation is currently available for this specific
combination. This underscores both the potential of the method in
identifying complex drug interactions and the need for further
validation through complementary studies.

As demonstrated in Section~\ref{sec-resFirth}, the post-treatment
logistic regression step provides clinical refinement by distinguishing
true interactions from combined individual effects.

This approach can also be extended to other settings where adverse
events are explored. For instance, the method could be applied using the
ICD diagnosis classification or the MedDRA system, both of which are
hierarchical classifications. Such an application would facilitate the
identification of symptoms associated with the consumption of drug
combinations.

The proposed method is implemented as an R package \emph{emcAdr},
available on GitHub and on the CRAN (Bangard 2025). A tool allowing the
processing of quarterly FAERS XML files to CSV files directly usable by
our method is also available on
\href{https://github.com/JulesBa-Git/FAERS-xml-parser}{GitHub}.

For researchers and stakeholders, it is crucial to remember that our
method is hypothesis-generating as are other signal detection methods
for single medications. These methods aim to generate as few
false-negative results as possible but with the trade-off of more false
positives. This means that any results should firstly be assessed for
biological plausibility as well as validated in a more formal causal
framework such as RCTs and target trial emulations. Furthermore, while
spontaneous reporting systems have a key role in pharmacovigilance, they
have known drawbacks as underlined in the introduction. They however
provide important knowledge about the safety of drugs, and with our
method also drug cocktails.

\section*{Funding}\label{funding}
\addcontentsline{toc}{section}{Funding}

This work was partially supported by the ``PHC AURORA'' programme
(project number: 49704QC), funded by the French Ministry for Europe and
Foreign Affairs, the French Ministry for Higher Education and Research
and the Norwegian Council for Research.

\section{Appendices}\label{appendices}

\subsection{Appendix A : Distance Pseudo-code Algorithm}\label{sec-appA}

\floatstyle{ruled}
\restylefloat{algorithm}
\begin{algorithm}[H]
\caption{\label{algo-distance}CalculateDistance}
\begin{algorithmic}[1]
\Require Matrix $M$, Lists $idxC_1$, $idxC_2$
\Ensure Cost of the distance calculation
\State $height \gets 0$
\State $cost \gets 0$
\While{$idxC_1 \neq \varnothing$ \textbf{and} $idxC_2 \neq \varnothing$}
  \ForAll{$iC_1 \in idxC_1$}
    \ForAll{$iC_2 \in idxC_2$}
      \If{$M[iC_1][height] = M[iC_2][height]$}
        \State $\begin{aligned}[t]
          cost \gets {}& cost + height \\
          &- \min\!\Big(M[iC_1].\mathrm{count}( M[iC_1][0]),\ \!M[iC_2].\mathrm{count}(M[iC_2][0])\Big) + 1
        \end{aligned}$
        \State \textbf{remove} $iC_1$ \textbf{from} $idxC_1$
        \State \textbf{remove} $iC_2$ \textbf{from} $idxC_2$
        \State \textbf{break}
      \EndIf
    \EndFor
  \EndFor
  \State $height \gets height + 1$
\EndWhile
\State $insertion\_cost \gets \mathrm{ATC\_HEIGHT} / 2$
\State $cost \gets cost + (|idxC_1| + |idxC_2|) \times insertion\_cost$
\State \Return $cost$
\end{algorithmic}
\end{algorithm}
\floatstyle{plain}

Algorithm~\ref{algo-distance} computes the distance between two
cocktails, \(C_1\) and \(C_2\), based on three inputs.

The first input is an integer matrix \(M\). This matrix represents the
nodes of \(C_1\) and \(C_2\), as well as their corresponding parent
nodes. The matrix has dimensions \((PN, \text{ATC\_HEIGHT})\), where
\(P\) and \(N\) denote the sizes of \(C_1\) and \(C_2\), respectively.
For example, consider the simplified tree shown in Figure 1 of the
article. In this tree, \(C_1\) could be the cocktail containing only the
node \(\begin{bmatrix} 3 \end{bmatrix}\), and \(C_2\) could be the
cocktail containing only the node \(\begin{bmatrix} 2 \end{bmatrix}\)
(\(P = 1\) and \(N = 1\)).

The corresponding matrix \(M\) would be: \[
M = \begin{bmatrix}
    3 & 1 & 0 \\
    2 & 2 & 0
\end{bmatrix}
\]

The second input consists of \(idxC_1\) and \(idxC_2\), which represent
the indices of the rows in \(M\) that contain the nodes of \(C_1\) and
\(C_2\), respectively.

\subsection{Appendix B : Calibration Under the Null}\label{sec-appB}

\begin{figure}

\centering{

\includegraphics[width=0.9\linewidth,height=\textheight,keepaspectratio]{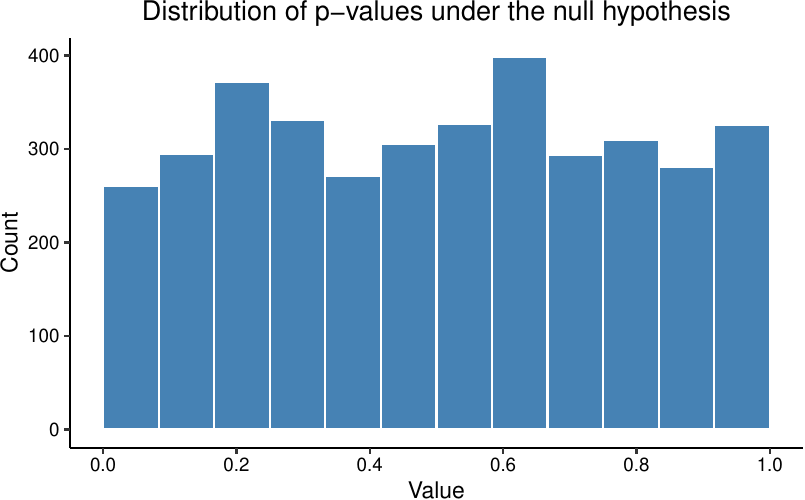}

}

\caption{\label{fig-calibration}Histogram of p-values for cocktails
presumed to satisfy the null hypothesis on Dataset \(D_1\) (no elevated
adverse-event risk).}

\end{figure}

To assess the validity of the approximation of the p-value, we plotted
the empirical distribution of p-values for cocktails presumed not to
increase the adverse-event rate on the Dataset \(D_1\). The theoretical
p-values following a Uniform(0, 1) distribution, the approximated ones
should exhibit a distribution close to uniform if the approximation is
valid. Figure~\ref{fig-calibration} shows a histogram of these p-values,
which appears approximately flat.

\subsection{Appendix C : FAERS
Clustering}\label{appendix-c-faers-clustering}

\begin{figure}

\centering{

\includegraphics[width=0.9\linewidth,height=\textheight,keepaspectratio]{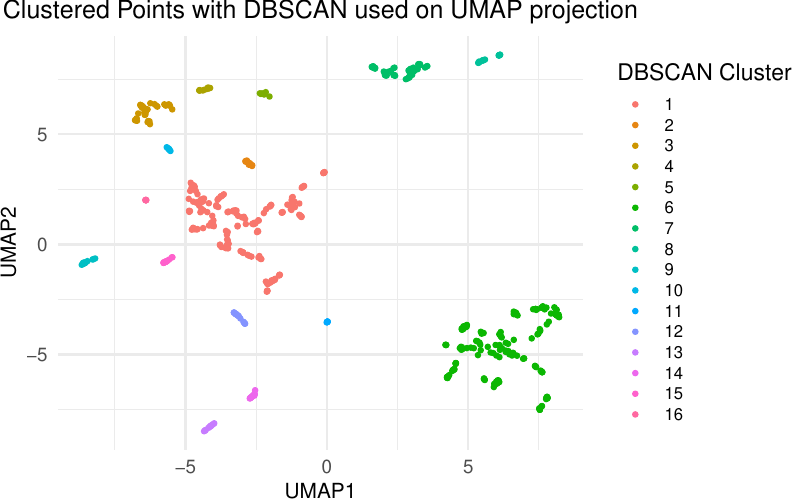}

}

\caption{\label{fig-cluster_FAERS}Clustering of High-Risk Drug Cocktails
Identified by the Genetic Algorithm on the FAERS dataset. This figure
allows one to see proximity between multiple clusters and better
understand why some clusters exhibit similarities.}

\end{figure}

Figure~\ref{fig-cluster_FAERS} shows the clustering applied to the
results of the genetic algorithm run on the FAERS datasets.

\subsection{Appendix D : Algorithmic
Complexity}\label{appendix-d-algorithmic-complexity}

The total time complexity of the high-risk cocktail identification is
given by: \[\mathcal{O}(G \cdot P \cdot T_{eval})\] Where \(G\)
represents the number of generations (or MCMC iterations) and \(P\)
represents the population size of the GA (this equals one for the MCMC
algorithm). The evaluation time for a single cocktail, \(T_{eval}\), is
defined by the matching logic required to count occurrences across the
database: \[T_{eval} = \mathcal{O}(N \cdot K \cdot R)\]

\begin{itemize}
\item
  \(N\) (Number of Reports): The algorithm iterates once through the
  database to compute the hypergeometric score for a given cocktail. The
  runtime scales linearly with the number of patients.
\item
  \(K\) (Cocktail Size): The number of drugs in the candidate cocktail.
  In this study, \(K\) is typically small since there is a limit to the
  number of drugs a patient takes.
\item
  \(R\) (Drugs per Report): The number of medications listed in a single
  patient report. On our data the mean size of drug cocktails per
  patient is approximately \(2.2\). Because \(K\) and \(R\) are bounded,
  the complexity simplifies to \(O(N)\) per candidate evaluation.
\end{itemize}

\subsection{Appendix E: Illustrative Example of Package
Usage}\label{appendix-e-illustrative-example-of-package-usage}

This appendix demonstrates the end-to-end usage of the \texttt{emcAdr}
package on a small simulated dataset that compiles in under a minute. It
is intended as a hands-on illustration of the API: full reproduction of
the results in Section~\ref{sec-appli_d1} and
Section~\ref{sec-appli-FAERS} requires the longer-running scripts
available in the article's GitHub repository. An accompanying
\texttt{README.md} file in the article's GitHub repository documents the
full reproduction workflow.

The toy dataset follows the same simulation design as Dataset \(D_1\)
(Section~\ref{sec-simu_data}), at a reduced scale and with deliberately
inflated adverse-event probabilities so that the signals remain
detectable in a small sample. Patients taking only one of the two
component drugs are added as decoys to break the spurious equivalence
between the size-two cocktail and either of its components in a small
dataset.

\subsubsection{E.1 Setup and Toy Dataset
Generation}\label{e.1-setup-and-toy-dataset-generation}

\begin{verbatim}
# A tibble: 7 x 3
  group       AE_FALSE AE_TRUE
  <chr>          <int>   <int>
1 C1_decoy_d1      132       0
2 C1_decoy_d2      134       2
3 C1_full           80      16
4 C2_decoy_d1      174       0
5 C2_decoy_d2      177       0
6 C2_full           84      10
7 random          4174      17
\end{verbatim}

The dataset contains 5000 patients, of whom approximately 2\% take the
full cocktail \(C_1\), 2\% the full cocktail \(C_2\), and 3\% take a
single drug from each cocktail as a decoy. The remaining patients
receive a random combination of active substances drawn from the ATC
tree leaves.

C1\_full patients show a \(20\)\% AE rate vs \(0.4\)\% in the random
group. Decoys show baseline AE rates as designed.

Please note that the columns of the patient dataframe have to be named
\texttt{PatientATC} and \texttt{PatientADR}.

\subsubsection{E.2 Evaluating Cocktail Risk via the Hypergeometric
Score}\label{e.2-evaluating-cocktail-risk-via-the-hypergeometric-score}

We start by computing the hypergeometric score \(H(C)\)
(Section~\ref{sec-risk-definition}) for a solutions cocktail and for a
random combination, to confirm that \(C_1\) is clearly separated on the
score scale.

\begin{verbatim}
[1] 37.79204  0.00000
\end{verbatim}

In datasets of this scale, the score function is sparse: most randomly
drawn size-2 pairs are taken by zero or one patient. A random pair
therefore has near \(100\)\% probability of scoring exactly zero, as
illustrated by random\_C above. The solution family \(C_1\) scores
\(H(C_1) \approx 38\), but whether this magnitude is extreme among
cocktails actually present in the data is the question answered by the
next section.

\subsubsection{E.3 Estimating the Null Distribution via
MCMC}\label{e.3-estimating-the-null-distribution-via-mcmc}

A short Metropolis-Hastings run (Section~\ref{sec-MCMC}) approximates
the distribution of \(H(C)\) for size-two cocktails. The full-scale
analysis uses substantially longer chains.

\begin{verbatim}
openMP available 
\end{verbatim}

\pandocbounded{\includegraphics[keepaspectratio]{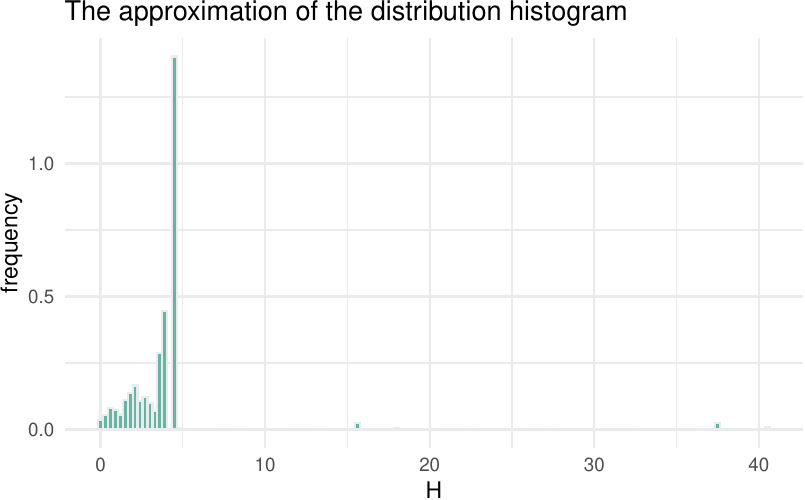}}

\begin{verbatim}
p-value of  solution C1 ( 47 1393 ) : 0.002672776
\end{verbatim}

The bulk of the distribution sits near zero. The visible spike around
\(H(C) \approx 4-5\) corresponds to a small-sample artifact: cocktails
taken by only one patient where the patient happens to experience the
AE. With only \(\sim 45\) AE cases over \(5000\) patients, such
configurations are common enough to form a visible noise peak.
Furthermore, with only \(5000\) data points, each cocktail tends to be
taken by few patients. The solution families generate \(H(C)\) values in
the \(18\)-\(40\) range, above this peak, which is why the empirical
p-value of \(C_1\) comes out at \(\sim 2.5 \times 10^{-3}\).

For size-two cocktails, computing the \emph{exact} score distribution is
tractable and offered by the package via
\texttt{trueDistributionSizeTwoCocktail()}. This allows assessing the
MCMC approximation directly (see also Figure~\ref{fig-distri_simu} in
the body):

The PP/QQ-plot comparison would confirm that the MCMC tracks the null
faithfully on the regions of \(H\) that actually occur in the data.

\subsubsection{E.4 Identifying High-Risk Cocktails with the Genetic
Algorithm}\label{e.4-identifying-high-risk-cocktails-with-the-genetic-algorithm}

To mirror the multi-run aggregation workflow of
Section~\ref{sec-appli_d1}, we run the genetic algorithm five times with
different seeds and pool the final populations. A single short run
typically converges on one of the two solutions. Aggregation across runs
increases the chance of recovering both. For full-scale runs on \(D_1\),
parallel execution scripts compatible with Slurm and similar workflow
managers are available in the article's GitHub repository under
\texttt{scripts/}.

\begin{verbatim}
# A tibble: 10 x 3
   cocktails score cocktails_display
   <list>    <dbl> <chr>            
 1 <int [2]>  40.6 0, 1393          
 2 <int [2]>  37.8 47, 1391         
 3 <int [2]>  37.6 45, 1393         
 4 <int [2]>  36.0 47, 1023         
 5 <int [1]>  33.2 1393             
 6 <int [2]>  18.9 630, 888         
 7 <int [2]>  18.9 658, 888         
 8 <int [2]>  18.7 659, 863         
 9 <int [2]>  18.6 659, 738         
10 <int [2]>  18.2 0, 888           
\end{verbatim}

The top outputs correspond to the solution families. Cocktails involving
the parent ATC nodes for \(C_1\) (nodes \(0\), \(45\), \(47\) paired
with \(1391\), \(1393\)) score in the \(36\)-\(40\) range. Those for
\(C_2\) (nodes \(630\), \(658\), \(659\) paired with \(888\), \(863\))
score around \(18\)-\(19\).

\subsubsection{E.5 Output Clustering}\label{e.5-output-clustering}

For didactic purposes we apply the same UMAP + DBSCAN pipeline as
described in Section~\ref{sec-clust} on the recovered cocktails. We
expect only a handful of distinct signals at this scale.

\pandocbounded{\includegraphics[keepaspectratio]{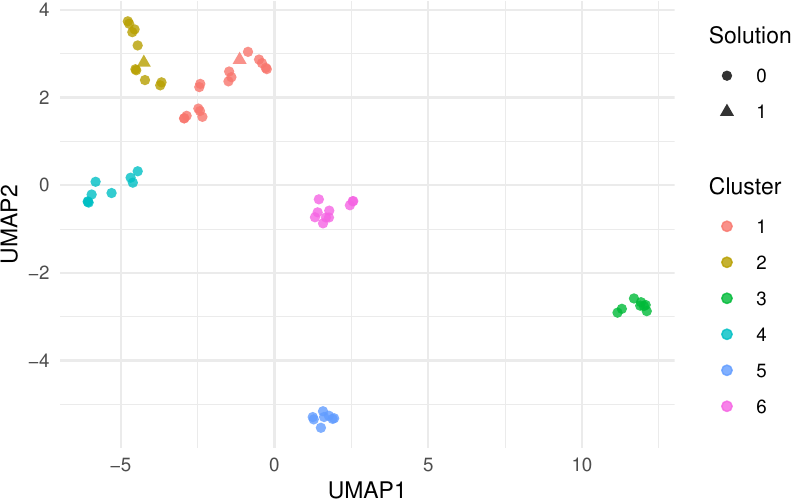}}

The two triangles (true solutions) embed in clusters 1 and 2 alongside
related GA cocktails, demonstrating that the dissimilarity-based
clustering groups variants of the same underlying signal in a reasonable
way.

\subsubsection{E.6 Post-Treatment with Firth's Penalized Logistic
Regression}\label{e.6-post-treatment-with-firths-penalized-logistic-regression}

For the \(C_1\) cocktails, we apply the post-treatment of
Section~\ref{sec-firth} to test whether the signal reflects a true
synergistic interaction or merely the additive effect of its components.
A function in \texttt{emcAdr} has been added in order to test a
cocktail.

\begin{verbatim}
logistf::logistf(formula = Y ~ ., data = df_model)

Model fitted by Penalized ML
Coefficients:
                   coef  se(coef) lower 0.95 upper 0.95     Chisq           p
(Intercept)   -5.188918 0.1985810 -5.6048123  -4.823127       Inf 0.000000000
X47.TRUE      -0.557285 1.4303230 -5.4020854   1.417979 0.1824861 0.669245382
X1393.TRUE     1.692411 0.5180643  0.5357799   2.615262 7.3118976 0.006849955
X47.1393.TRUE  2.468896 1.5322560  0.1735472   7.390559 4.6094408 0.031796401
              method
(Intercept)        2
X47.TRUE           2
X1393.TRUE         2
X47.1393.TRUE      2

Method: 1-Wald, 2-Profile penalized log-likelihood, 3-None

Likelihood ratio test=80.73281 on 3 df, p=0, n=5000
Wald test = 787.0257 on 3 df, p = 0
\end{verbatim}

The p-value of \(0.032\) on the interaction term confirms that this
synergistic effect is statistically distinguishable, even with only 16
AE cases in the \texttt{C1\_full} group. It confirms that the genetic
algorithm's discovery reflects a true drug--drug interaction rather than
the marginal contribution of either family alone.

Genetic algorithm output also contains cocktail \(1393\) which
contributes significantly to the AE on its own as depicted in the
summary.

\section{Code and Results}\label{code-and-results}

See the complete results of application of methodology on the FAERS
dataset
\href{https://docs.google.com/spreadsheets/d/1TmbrExi26QNr2TmvltmEzyl_EQht3fIsrtsWrAqmRik/edit?usp=sharing}{on
this link}.

Code to reproduce the experiments conducted in this article can be found
here :

\begin{itemize}
\tightlist
\item
  GitHub link of the data refinement code :
  \href{https://github.com/JulesBa-Git/FAERS-xml-parser}{JulesBa-Git/FAERS-xml-parser}.
\item
  GitHub link to the latest release of the R package of the complete
  methodology :
  \href{https://github.com/JulesBa-Git/emcAdr}{JulesBa-Git/emcAdr}.
\item
  The code of the methodology is also available on the
  \href{https://cran.r-project.org/web/packages/emcAdr/index.html}{CRAN}.
\item
  The code to obtain figures is available in the \texttt{scripts} folder
  of this repository.
\end{itemize}

\section*{References}\label{references}
\addcontentsline{toc}{section}{References}

\phantomsection\label{refs}

\section*{Session information}\label{session-information}
\addcontentsline{toc}{section}{Session information}

\begin{Routput}
R version 4.5.1 (2025-06-13)
Platform: aarch64-apple-darwin20
Running under: macOS Tahoe 26.4.1

Matrix products: default
BLAS:   /Library/Frameworks/R.framework/Versions/4.5-arm64/Resources/lib/libRblas.0.dylib 
LAPACK: /Library/Frameworks/R.framework/Versions/4.5-arm64/Resources/lib/libRlapack.dylib;  LAPACK version 3.12.1

locale:
[1] fr_FR/UTF-8/fr_FR/C/fr_FR/fr_FR

time zone: Europe/Paris
tzcode source: internal

attached base packages:
[1] stats     graphics  grDevices utils     datasets  methods   base     

other attached packages:
[1] tidyr_1.3.1   purrr_1.1.0   stringr_1.5.1 umap_0.2.10.0 ggplot2_3.5.2
[6] dbscan_1.2.2  dplyr_1.1.4   emcAdr_1.3   

loaded via a namespace (and not attached):
 [1] gtable_0.3.6         shape_1.4.6.1        xfun_0.52           
 [4] formula.tools_1.7.1  lattice_0.22-7       vctrs_0.6.5         
 [7] tools_4.5.1          Rdpack_2.6.4         generics_0.1.4      
[10] tibble_3.3.0         pan_1.9              pkgconfig_2.0.3     
[13] jomo_2.7-6           Matrix_1.7-3         RColorBrewer_1.1-3  
[16] lifecycle_1.0.4      compiler_4.5.1       farver_2.1.2        
[19] tinytex_0.57         codetools_0.2-20     htmltools_0.5.8.1   
[22] yaml_2.3.10          glmnet_4.1-10        mice_3.19.0         
[25] nloptr_2.2.1         pillar_1.11.0        MASS_7.3-65         
[28] openssl_2.3.3        reformulas_0.4.4     iterators_1.0.14    
[31] rpart_4.1.24         boot_1.3-31          foreach_1.5.2       
[34] mitml_0.4-5          nlme_3.1-168         RSpectra_0.16-2     
[37] tidyselect_1.2.1     digest_0.6.37        stringi_1.8.7       
[40] labeling_0.4.3       splines_4.5.1        operator.tools_1.6.3
[43] fastmap_1.2.0        grid_4.5.1           cli_3.6.5           
[46] magrittr_2.0.3       survival_3.8-3       utf8_1.2.6          
[49] broom_1.0.9          withr_3.0.2          scales_1.4.0        
[52] backports_1.5.0      logistf_1.26.1       rmarkdown_2.29      
[55] nnet_7.3-20          lme4_2.0-1           reticulate_1.43.0   
[58] askpass_1.2.1        png_0.1-8            evaluate_1.0.4      
[61] knitr_1.50           rbibutils_2.3        mgcv_1.9-3          
[64] rlang_1.1.6          Rcpp_1.1.0           glue_1.8.0          
[67] minqa_1.2.8          jsonlite_2.0.0       R6_2.6.1            
\end{Routput}

\end{document}